\documentclass[twocolumn]{aastex631}
\usepackage[normalem]{ulem}  
\usepackage[T1]{fontenc}
\usepackage{ae,aecompl}
\usepackage{graphicx}
\usepackage{amsmath}
\usepackage{amssymb}
\usepackage{supertabular}
\usepackage{hyperref}
\usepackage{xcolor}
\usepackage{academicons}
\usepackage{multirow}
\usepackage{booktabs}
\usepackage[normalem]{ulem}

\begin{document}
\title{Neutron Star Mass-Radius Constraints for EXO~0748$-$676 from 2008-2025 Quiescent X-ray Spectra 
}
\author{Mingyang Wang}
\affiliation{Department of Astronomy, Xiamen University, Xiamen 361005, P.R. China; liang@xmu.eud.cn}

\author{Guobao Zhang}
\affiliation{Yunnan Observatories, Chinese Academy of Sciences (CAS), Kunming 650216, P.R. China}
\affiliation{Key Laboratory for the Structure and Evolution of Celestial Objects, CAS, Kunming 650216, P.R. China}

\author{Ang Li}
\affiliation{Department of Astronomy, Xiamen University, Xiamen 361005, P.R. China; liang@xmu.eud.cn}

\begin{abstract}

We present new constraints on the mass and radius of the neutron star in the neutron star low-mass X-ray binary EXO 0748$-$676 obtained from a joint analysis of 20 quiescent X-ray observations obtained between 2008 and 2025, including 14 \textit{Chandra} and 6 \textit{XMM-Newton} exposures. These data sample two quiescent episodes separated by the 2024$-$2025 outburst. 
We model the 0.5--10~keV spectra with a hydrogen-atmosphere model, assuming a source distance of 7.1~kpc. In a global Markov Chain Monte Carlo analysis in which the hydrogen column density, neutron star mass, and radius are tied across all observations, we obtain a neutron-star mass of $1.77^{+0.17}_{-0.22}\,M_\odot$ and a radius of $12.62^{+0.56}_{-0.74}$ km ($1\sigma$ credible intervals). We further perform independent fits to the first and second quiescent epochs and find that the combined data set significantly reduces the low-mass tail in the posterior distribution, leading to tighter lower bounds on the neutron-star mass.
Incorporating the distance uncertainty of $7.1\pm1.2$~kpc, we conservatively constrain the neutron-star mass and radius to $M\simeq 1.41-2.11~M_{\odot}$ and $R\simeq 10.15-15.13$ km, favoring relatively stiff dense-matter equations of state. We also trace the thermal evolution across two quiescent epochs and find evidence for renewed crust cooling following the 2024$-$2025 outburst, providing a unique opportunity to compare the thermal relaxation behavior after two distinct accretion episodes.
 
\end{abstract}


\section{Introduction}
\label{sec:intro}

Neutron stars represent the densest observable matter in the Universe, with central densities reaching several times nuclear saturation density ($\rho_0 \approx 2.8 \times 10^{14}$~g~cm$^{-3}$). The relationship between pressure and density in this regime---the equation of state (EOS) of cold, catalyzed matter---remains one of the most important open problems in nuclear astrophysics \citep[e.g.,][]{2016RvMP...88b1001W, 2022NatRP...4..237Y}. Because the EOS directly determines the mass--radius ($M$--$R$) relation of neutron stars, precise measurements of neutron-star masses and radii provide powerful constraints on the behavior of matter at densities far exceeding those achievable in terrestrial laboratories \citep[e.g.,][]{2025SCPMA..6819503L}.

EXO~0748$-$676  (also known as UY~Vol) is a neutron star low-mass X-ray binary (NS LMXB) that has served as one of the most important laboratories for constraining the fundamental properties of neutron stars. Since its discovery in 1985 \citep{1985IAUC.4039....1P}, this system has provided multiple independent pathways to measure the neutron star mass and radius, including spectroscopic measurements of gravitational redshift~\citep{2002Natur.420...51C, 2006Natur.441.1115O} 
observations of thermonuclear burst oscillations~\citep{2004ApJ...614L.121V, 2010ApJ...711L.148G}, 
quiescent thermal emission analysis~\citep{2011MNRAS.414.1077Z, 2011AA...528A.150D, 2014ApJ...791...47D, 2017MNRAS.471.2605C}, 
and dynamical studies of the binary system~\citep{2009MNRAS.394L.136M, 2022MNRAS.510.4736K}.

Among the various observational approaches to measuring neutron star properties, the analysis of thermal emission from quiescent low-mass X-ray binaries (qLMXBs) has proven particularly fruitful \citep{1998A&ARv...8..279C,2008AIPC..983..519J}. 
In these systems, accretion onto the neutron-star surface ceases or diminishes dramatically during quiescent episodes, allowing the thermal glow of the neutron-star crust and core to be observed directly \citep{2004cxo..prop.1673W}. 
Because mass transfer from the low-mass companion is generally hydrogen rich, gravitational settling rapidly drives the lightest elements to the surface, producing a hydrogen-dominated atmosphere on timescales much shorter than the outburst–quiescence cycle. 
Consequently, the observed soft X-ray spectra can be modeled with realistic neutron-star hydrogen-atmosphere models to infer constraints on the stellar mass and radius \citep{2006ApJ...644.1090H,2002nsps.conf..263Z}.

EXO 0748$-$676 occupies a particularly important place in this context. Discovered as a bright X-ray transient with the \textit{EXOSAT} satellite \citep{1999AJ....118.1390G}, it is an eclipsing LMXB with an orbital period of 3.82 hr \citep{1986ApJ...308..213G}. The system exhibits deep X-ray eclipses of $\approx 8$ min duration and periodic absorption dips attributed to structure in the accretion disk, implying a high inclination of $i \approx 75^{\circ}$--$80^{\circ}$ \citep{2022MNRAS.510.4736K}. The donor is a late-type star with $M_2 \approx 0.4$--$0.6M_{\odot}$ \citep{2009MNRAS.394L.136M}. 

Previously, EXO~0748$-$676 underwent a quasi-persistent outburst lasting more than 24 years before returning to quiescence in late 2008 \citep{2009MNRAS.396L..26D, 2011MNRAS.412.1409D}. Subsequent long-term monitoring in quiescence with \textit{Chandra}, \textit{XMM-Newton}, \textit{Swift}, and the \textit{Hubble Space Telescope} has provided an exceptionally rich dataset for studying neutron-star thermal emission, crust cooling, and mass--radius constraints~\citep{2014ApJ...791...47D, 2017MNRAS.471.2605C,2020A&A...638L...2P,parikh2021}.

After about 16 years in quiescence, EXO 0748$-$676 returned to outburst in early 2024~\citep{2024GCN.36653....1D}. The reactivation was detected by \textit{Swift}/BAT on 2024 June 10, triggering extensive
follow-up observations across the electromagnetic spectrum~\citep[e.g.,][]{2024ApJ...977L..17B, 2024arXiv241006201S, 2026JHEAp..5100535A, 2026JHEAp..5300595S}.

In this work, we present a comprehensive analysis of all available quiescent X-ray observations of EXO 0748$-$676 obtained with \textit{Chandra} and \textit{XMM-Newton} between 2008 and 2025, encompassing two quiescent episodes separated by $\approx16$ years. Our goals are to (i) derive updated mass-radius constraints from a global spectral analysis that self-consistently combines all available data, (ii) quantify the impact of distance uncertainties on the inferred neutron-star parameters, and (iii) place these results in the context of previous constraints on EXO 0748$-$676 and current dense-matter EOS models. 
The structure of the paper is as follows. Section 2 describes the observations, data reduction, and spectral modeling, and presents the MCMC framework used to obtain mass–radius constraints. Section 3 presents the resulting constraints and explores their dependence on data subsets and distance. Section 4 discusses the implications for neutron-star structure and the dense-matter EOS, and compares our results with previous work. Finally, Section 5 summarizes our main conclusions and outlines prospects for future observations.

\begin{table*}[ht]
\centering
\caption{
\textit{Chandra} and \textit{XMM-Newton} observations of EXO 0748$-$676 used in this work. 
The last three \textit{Chandra} observations (ObsIDs 30047--30049) form the second quiescent epoch shown in red in Figure~\ref{fig:Teff}.
}
\label{tab:obs}
\renewcommand{\arraystretch}{1.2}
\setlength{\tabcolsep}{8pt}
\begin{tabular}{ccccccc}
\hline\hline
Instr. & ObsID & Date & MJD & Detector & Exposure (ks) & Quiescent epoch \\
\hline

\textit{Chandra} & 9070  & 2008-10-12 & 54751 & ACIS-S & 13.77 & \multirow{17}{*}{1} \\
\textit{Chandra} & 10783 & 2008-10-15 & 54754 & ACIS-S & 13.29 & \\
\textit{XMM-Newton} & 0560180701 & 2008-11-06 & 54776 & MOS1/MOS2/PN & 29.52/29.52/27.93 & \\

\textit{Chandra} & 9071  & 2009-02-23 & 54885 & ACIS-S & 15.83 & \\
\textit{Chandra} & 10871 & 2009-02-25 & 54887 & ACIS-S & 9.58 & \\
\textit{XMM-Newton} & 0605560401 & 2009-03-18 & 54908 & MOS1/MOS2/PN & 43.82/42.67/41.91 & \\

\textit{Chandra} & 9072  & 2009-06-10 & 54992 & ACIS-S & 27.22 & \\
\textit{XMM-Newton} & 0605560501 & 2009-07-01 & 55013 & MOS1/MOS2/PN & 101.62/101.62/100.03 & \\

\textit{Chandra} & 11059 & 2010-04-20 & 55306 & ACIS-S & 27.37 & \\
\textit{XMM-Newton} & 0651690101 & 2010-06-17 & 55364 & MOS1/MOS2/PN & 98.87/98.89/96.40 & \\

\textit{Chandra} & 11060 & 2010-10-20 & 55489 & ACIS-S & 27.22 & \\
\textit{Chandra} & 12414 & 2011-07-02 & 55744 & ACIS-S & 38.07 & \\

\textit{XMM-Newton} & 0690330101 & 2013-04-15 & 56397 & MOS1/MOS2/PN & 104.07/104.12/102.54 & \\
\textit{Chandra} & 14663 & 2013-08-01 & 56505 & ACIS-S & 42.86 & \\

\textit{XMM-Newton} & 0824420101 & 2018-05-02 & 58240 & MOS1/MOS2/PN & 77.62/77.60/75.76 & \\

\textit{Chandra} & 23440 & 2021-08-03 & 59429 & ACIS-S & 26.83 & \\
\textit{Chandra} & 24516 & 2021-09-21 & 59478 & ACIS-S & 23.60 & \\
\hline
\textit{Chandra} & 30047 & 2025-05-23 & 60818 & ACIS-S & 14.14 & \multirow{3}{*}{2} \\
\textit{Chandra} & 30048 & 2025-06-17 & 60843 & ACIS-S & 18.43 & \\
\textit{Chandra} & 30049 & 2025-08-11 & 60898 & ACIS-S & 23.50 & \\
\hline
\end{tabular}
\end{table*}

\section{Observations and Data Analysis}
\label{sec:quiescence}

\subsection{Observations}

The basic properties of the 20 observations analyzed in this work are listed in Table \ref{tab:obs}. The sample includes 14 \textit{Chandra}/ACIS-S observations and 6 \textit{XMM-Newton} observations with EPIC MOS1, MOS2, and PN detectors, obtained between 2008 and 2025. All exposures were taken when the source was in quiescence, following either the long 1985$-$2008 outburst~\citep{2009MNRAS.396L..26D} or the renewed 2024$-$2025 outburst~\citep{2025ATel17191....1D}. 

\subsection{Data Reduction}

\textit{Chandra} data were reduced using CIAO v4.18 \citep{2006SPIE.6270E..1VF}. We first reprocessed the level-1 event files with \texttt{chandra\_repro} using CALDB v4.12.3. We then inspected the light curves of the level-2 event files with \textsc{xselect}, using the same source region as that adopted for the spectral extraction, and inspected them for possible eclipses. We removed the eclipse intervals based on the known eclipse period and duration of the source \citep{1993AAS...183.5515H} as well as intervals affected by possible background flares. Source spectra were extracted from circular regions of radius $\sim3''$ centered on EXO 0748$-$676, and background spectra were extracted from source-free annular regions with inner and outer radii of $\sim10''$ and $\sim25''$, respectively, excluding any nearby contaminating sources. The corresponding response matrix files (RMFs) and ancillary response files (ARFs) were generated with the \textsc{specextract} tool.

\textit{XMM-Newton} data were reduced using SAS v22.1.0. The PN level-2 event files were generated with \texttt{epchain}, and the MOS1 and MOS2 level-2 event files were generated with \texttt{emchain}. Eclipse intervals and possible background flares were also removed according to the light curves, after which the spectra were extracted with \texttt{especget}. Source spectra were extracted from circular regions of radius $\sim30''$ centered on the source, while background spectra were extracted from nearby source-free regions on the same CCD. In the subsequent spectral fitting, we tied all parameters of the MOS1, MOS2, and PN spectra from the same observation.

After spectral extraction, the spectra were grouped with \texttt{grppha} to ensure a minimum of 25 counts per bin. Spectral fitting was performed using \textsc{XSPEC} v12.15.1 \citep{1996ASPC..101...17A}. 

\begin{table*}
\centering
\small
\setlength{\tabcolsep}{4.5pt}
\renewcommand{\arraystretch}{1.15}
\caption{Spectral-fit results ordered by observation date. The hydrogen column density, neutron-star mass, and neutron-star radius were tied across all observations; 
Their fitted values are $N_{\rm H} = 0.58\pm 0.03\times 10^{21}\,\mathrm{cm}^{-2}$, $M = 1.77^{+0.17}_{-0.22}\,M_\odot$, $R = 12.62^{+0.56}_{-0.74}\,\mathrm{km}$. 
The power-law photon index was tied within each quiescent epoch but allowed to vary between epochs. The last column lists the fit statistic $\chi^2$ and the corresponding number of degrees of freedom (dof) for each dataset. The values of $\chi^2$ and dof indicate that most datasets are fitted reasonably well by the adopted model.
}
\label{tab:spectral_fit_summary}
\renewcommand{\arraystretch}{1.2}
\setlength{\tabcolsep}{4.5pt}
\begin{tabular}{ccccccccc}
\hline \hline
Instr. & ObsID & Date & $kT_{\rm eff}$ & $\Gamma$ & $K_{\rm PL}$ & PL frac. & $F_{0.5-10\,\mathrm{keV}}$ & $\chi^2/\mathrm{dof}$ \\
 &  &  & $\mathrm{keV}$ &  & \shortstack{$10^{-5}$\\$\mathrm{ph\,keV^{-1}\,cm^{-2}\,s^{-1}}$} & $\%$ & \shortstack{$10^{-12}$\\$\mathrm{erg\,cm^{-2}\,s^{-1}}$} &  \\
\hline
\textit{Chandra} & 9070 & 2008-10-12 & $0.160^{+0.010}_{-0.009}$ & \multirow{17}{*}{$1.86^{+0.12}_{-0.13}$} & $4.25^{+1.11}_{-1.01}$ & $23.9^{+5.2}_{-4.8}$ & $1.252^{+0.038}_{-0.037}$ & $94.48/80$ \\
\textit{Chandra} & 10783 & 2008-10-15 & $0.162^{+0.010}_{-0.009}$ &  & $3.70^{+1.12}_{-0.97}$ & $20.4^{+5.3}_{-4.8}$ & $1.253^{+0.038}_{-0.038}$ & $79.22/78$ \\
\textit{XMM-Newton} & 0560180701 & 2008-11-06 & $0.160^{+0.010}_{-0.009}$ &  & $1.52^{+0.36}_{-0.31}$ & $10.1^{+1.5}_{-1.5}$ & $1.069^{+0.013}_{-0.013}$ & $187.90/157$ \\
\textit{Chandra} & 9071 & 2009-02-23 & $0.154^{+0.010}_{-0.009}$ &  & $2.09^{+1.33}_{-1.18}$ & $18.5^{+6.6}_{-6.0}$ & $0.941^{+0.031}_{-0.030}$ & $76.47/73$ \\
\textit{Chandra} & 10871 & 2009-02-25 & $0.152^{+0.010}_{-0.009}$ &  & $3.71^{+2.42}_{-2.12}$ & $26.9^{+9.0}_{-8.1}$ & $1.022^{+0.042}_{-0.041}$ & $45.56/47$ \\
\textit{XMM-Newton} & 0605560401 & 2009-03-18 & $0.153^{+0.010}_{-0.009}$ &  & $0.05^{+0.07}_{-0.03}$ & $<1\%$ & $0.804^{+0.009}_{-0.009}$ & $203.42/155$ \\
\textit{Chandra} & 9072 & 2009-06-10 & $0.150^{+0.010}_{-0.009}$ &  & $1.00^{+0.61}_{-0.55}$ & $11.4^{+4.9}_{-4.5}$ & $0.794^{+0.021}_{-0.020}$ & $98.22/87$ \\
\textit{XMM-Newton} & 0605560501 & 2009-07-01 & $0.148^{+0.010}_{-0.008}$ &  & $0.22^{+0.12}_{-0.10}$ & $2.8^{+1.0}_{-1.0}$ & $0.695^{+0.005}_{-0.005}$ & $253.05/221$ \\
\textit{Chandra} & 11059 & 2010-04-20 & $0.147^{+0.010}_{-0.008}$ &  & $1.83^{+0.68}_{-0.60}$ & $17.0^{+4.9}_{-4.5}$ & $0.787^{+0.022}_{-0.021}$ & $75.42/86$ \\
\textit{XMM-Newton} & 0651690101 & 2010-06-17 & $0.148^{+0.010}_{-0.008}$ &  & $0.03^{+0.04}_{-0.02}$ & $<1\%$ & $0.684^{+0.006}_{-0.005}$ & $394.12/364$ \\
\textit{Chandra} & 11060 & 2010-10-20 & $0.146^{+0.010}_{-0.009}$ &  & $1.64^{+0.58}_{-0.53}$ & $15.8^{+4.8}_{-4.3}$ & $0.745^{+0.021}_{-0.020}$ & $70.42/84$ \\
\textit{Chandra} & 12414 & 2011-07-02 & $0.147^{+0.009}_{-0.008}$ &  & $2.31^{+0.52}_{-0.46}$ & $19.7^{+3.5}_{-3.2}$ & $0.800^{+0.018}_{-0.018}$ & $83.70/101$ \\
\textit{XMM-Newton} & 0690330101 & 2013-04-15 & $0.142^{+0.009}_{-0.008}$ &  & $0.06^{+0.06}_{-0.04}$ & $1.1^{+0.9}_{-0.9}$ & $0.569^{+0.005}_{-0.005}$ & $255.08/175$ \\
\textit{Chandra} & 14663 & 2013-08-01 & $0.138^{+0.009}_{-0.008}$ &  & $1.06^{+0.42}_{-0.38}$ & $15.6^{+4.8}_{-4.4}$ & $0.572^{+0.016}_{-0.015}$ & $96.71/87$ \\
\textit{XMM-Newton} & 0824420101 & 2018-05-02 & $0.146^{+0.009}_{-0.008}$ &  & $0.61^{+0.14}_{-0.13}$ & $6.2^{+1.1}_{-1.0}$ & $0.671^{+0.006}_{-0.006}$ & $200.33/175$ \\
\textit{Chandra} & 23440 & 2021-08-03 & $0.148^{+0.010}_{-0.009}$ &  & $1.67^{+0.49}_{-0.42}$ & $13.1^{+4.4}_{-3.7}$ & $0.776^{+0.037}_{-0.036}$ & $27.24/40$ \\
\textit{Chandra} & 24516 & 2021-09-21 & $0.148^{+0.010}_{-0.009}$ &  & $0.81^{+0.55}_{-0.46}$ & $7.5^{+5.7}_{-4.7}$ & $0.742^{+0.038}_{-0.037}$ & $20.67/33$ \\
\hline
\textit{Chandra} & 30047 & 2025-05-23 & $0.147^{+0.010}_{-0.009}$ & \multirow{3}{*}{$2.4^{+0.20}_{-0.32}$} & $4.58^{+2.60}_{-2.44}$ & $23.7^{+10.7}_{-8.8}$ & $0.853^{+0.067}_{-0.064}$ & $16.20/17$ \\
\textit{Chandra} & 30048 & 2025-06-17 & $0.146^{+0.010}_{-0.009}$ &  & $5.95^{+1.93}_{-2.24}$ & $28.2^{+8.3}_{-7.0}$ & $0.881^{+0.057}_{-0.055}$ & $25.15/24$ \\
\textit{Chandra} & 30049 & 2025-08-11 & $0.142^{+0.010}_{-0.008}$ &  & $5.70^{+1.93}_{-2.15}$ & $29.6^{+7.1}_{-6.1}$ & $0.797^{+0.048}_{-0.046}$ & $31.66/27$ \\
\hline \hline
\end{tabular}
\end{table*}

\subsection{Spectral Analysis}

Transient LMXBs undergo alternating periods of active accretion (outbursts) and quiescence. During outbursts, compressional heating of the neutron-star crust by accreted matter drives the crust out of thermal equilibrium with the core \citep{2004cxo..prop.1673W, 2012arXiv1201.5602P}. When accretion ceases and the system enters quiescence, the crust cools by thermal radiation, producing a soft X-ray spectrum that can be modeled as thermal emission from the neutron-star surface \citep{1998A&ARv...8..279C}.

The observed quiescent X-ray spectrum is typically well described by a combination of a soft thermal component, modeled with a neutron-star atmosphere spectrum, and a harder power-law tail that may arise from residual accretion or other non-thermal processes~\citep{2008AIPC..983..519J,2018MNRAS.479.3634M}.
The thermal component is characterized by an effective temperature $T_{\rm eff}$ and is modified by interstellar absorption, quantified by the column density $N_{\rm H}$.
The neutron-star atmosphere models should incorporate the effects of the extreme surface gravity ($g \sim 10^{14}$~cm~s$^{-2}$), the composition of the atmosphere (typically hydrogen or helium), and the possibly strong magnetic fields~\citep{2002nsps.conf..263Z,2009A&A...500..891S}.
The emergent spectrum differs significantly from a simple blackbody due to electron scattering and free-free absorption in the atmosphere. Fitting the observed spectrum with such models yields constraints on $M$ and $R$ for an assumed distance $d$~\citep{2006ApJ...644.1090H,2014ApJ...784..123L}. 

The quiescent X-ray spectra of EXO~0748$-$676 have been modeled using several approaches. The thermal component is typically described by the \texttt{nsatmos} model \citep{2006ApJ...644.1090H}, which assumes a non-magnetized hydrogen atmosphere in hydrostatic and radiative equilibrium. 
No persistent accretion-powered pulsations have been detected from EXO~0748$-$676, consistent with the weak magnetic fields generally inferred for neutron stars in LMXBs and supporting the use of non-magnetic atmosphere models. 
This model has three free parameters: the neutron-star mass $M$, the radius $R$, and the effective temperature $kT_{\rm eff}$, in addition to a normalization which depends on the source distance $d$.
Alternatively, the \texttt{nsagrav} model \citep{2002nsps.conf..263Z}, which uses a fully ionized hydrogen atmosphere with self-consistent treatment of radiative transfer, has been employed by \citet{2011MNRAS.414.1077Z} and \citet{2017MNRAS.471.2605C}. Both models yield qualitatively similar constraints, though the exact best-fit values depend on the model assumptions. 
Based on our tests on all \textit{Chandra} and \textit{XMM-Newton} observations between 2008 and 2009, we find that the \texttt{nsagrav} model yields a mass approximately 5.5\% higher and a radius approximately 3.8\% lower than those obtained with the \texttt{nsatmos} model. 

An additional hard component, typically modeled as a power law, is often included to account for non-thermal emission. \citet{2011MNRAS.412.1409D} found that this component contributes approximately 4\%--20\% of the total unabsorbed 0.5--10~keV flux, with a photon index of $\Gamma \sim 1$--2.

Interstellar absorption is modeled with the \texttt{tbabs} model \citep{wilms2000}, with the column density $N_{\rm H}$ either fixed to the Galactic value or allowed to vary. The fitted $N_{\rm H}$ values for EXO~0748$-$676 are typically in the range $\sim(0.7$--$1.2) \times 10^{21}$~cm$^{-2}$ \citep{2011MNRAS.412.1409D}, consistent with the expected extinction toward the source.

For our analysis, we adopted the model \texttt{tbabs*(nsatmos+powerlaw)} \citep{2006ApJ...644.1090H}, with the interstellar absorption modeled using the \cite{wilms2000} abundance set and the \cite{1996ApJ...465..487V} photoelectric cross sections.

Current distance constraints for EXO~0748$-$676 are centered around $\sim 7\,\mathrm{kpc}$ \citep[e.g.,][]{2005ApJ...632.1099W, 2008MNRAS.387..268G, 2011MNRAS.414.1077Z, 2026JHEAp..5100535A}, but with considerable uncertainty. We therefore fixed the source distance at 
$7.1\,\mathrm{kpc}$ in our baseline analysis, and later quantified the impact of the distance uncertainty over a wide range from $5.9\,\mathrm{kpc}$ to $8.3\,\mathrm{kpc}$ \citep{2008MNRAS.387..268G}. The \texttt{nsatmos} normalization was fixed at unity, corresponding to emission from the entire neutron-star surface, while $N_{\rm H}$, $M$, and $R$ were linked across all observations.
Considering that the non-thermal emission of the source may differ between the two quiescent epochs, we tied the power-law photon index within each epoch but allowed it to vary between epochs. 

To verify the validity of our assumption of tying the photon index within each quiescent epoch, we further divided the data from the first quiescent epoch into three segments and performed independent spectral fits to obtain the photon index for each segment, in order to examine whether the photon index exhibits significant variations (see Section~3.2).
The results show that the photon index remains broadly consistent within uncertainties, supporting our assumption of tying it within each epoch.

The effective temperature ($T_{\rm eff}$) and the power-law normalization at 1 $\mathrm{keV}$ ($K_{\rm PL}$) were allowed to vary among all observations.
The spectral fitting was performed in the 0.5--10.0 keV energy range. 
Given the reported energy dependence of spectral fitting results \citep{2017MNRAS.471.2605C}, we performed additional tests using the 2008--2009 data. Fitting in the 0.5--10.0\,keV band yielded \(M=1.63^{+0.26}_{-0.37}\,M_\odot\) and \(R=12.94^{+0.69}_{-0.94}\,\mathrm{km}\); extending the band to 0.3--10.0\,keV gave \(M=1.89^{+0.09}_{-0.15}\,M_\odot\) and \(R=12.26^{+0.64}_{-0.63}\,\mathrm{km}\). These results indicate that the fits at low energies are affected by multiple factors. First, low photon statistics play a role; \citet{2018MNRAS.479.3634M} showed through simulations that the energy dependence can be eliminated with substantially improved data quality. Second, \citet{2017MNRAS.471.2605C} found significant residuals near 0.5\,keV, which they attributed to a thermal gas component that contaminates the neutron-star atmospheric emission. In addition, low-energy bands are generally more susceptible to absorption and calibration uncertainties. Given these various uncertainties at low energies, we adopted a relatively conservative energy range of 0.5--10.0\,keV for our analysis.

\subsection{MCMC Setup and Parameter Inference}

Given the comprehensive set of quiescent spectra and the relatively large number of free parameters in this work, we used the Markov chain Monte Carlo (MCMC) method to more thoroughly explore the model parameter space and obtain the posterior distributions of all parameters, thereby providing a more robust assessment of the uncertainties in the neutron-star mass and radius. 
Specifically, we first used conventional best-fit optimization to determine the approximate ranges and orders of magnitude of the model parameters. The allowed parameter ranges were set to \(N_{\rm H}=0.01\)--\(0.20\times10^{22}\,\mathrm{cm^{-2}}\), \(\log(T_{\rm eff}/{\rm K})=6.0\)--6.5, \(M=1.0\)--\(2.8\,M_\odot\), \(R=5\)--20\,km, photon index \(\Gamma=0.5\)--3.0, and power-law normalization \(K_{\rm PL}=0\)--\(1\times10^{-4}\,\mathrm{ph\,keV^{-1}\,cm^{-2}\,s^{-1}}\). We then performed MCMC sampling with the Goodman--Weare algorithm and initialized the walkers based on the best-fit solution and the associated covariance matrix. To reduce the risk of the chains becoming trapped in local minima, we further applied random offsets of about 10\%--30\% to the best-fit parameters and ran multiple chains in parallel from different initial values. We used 300 walkers and evolved the chains for a total of \(4\times10^{7}\) steps, with a typical burn-in length of \(1.5\times10^{7}\) steps. We assessed convergence using the criterion $\hat{R}<1.1$ and discarded the initial burn-in segment of each chain in the analysis. In addition, we also tested initializing the walkers with a uniform distribution within the allowed parameter ranges; this yielded posterior results consistent with those obtained from Gaussian initialization, but the convergence was significantly slower.
The final posterior distributions were obtained for $N_{\rm H}$, the neutron-star mass and radius, the photon indices for the two quiescent epochs, as well as the 20 effective temperatures and power-law normalizations.

\section{RESULTS}

\begin{figure}
\centering
\includegraphics[width=0.45\textwidth]{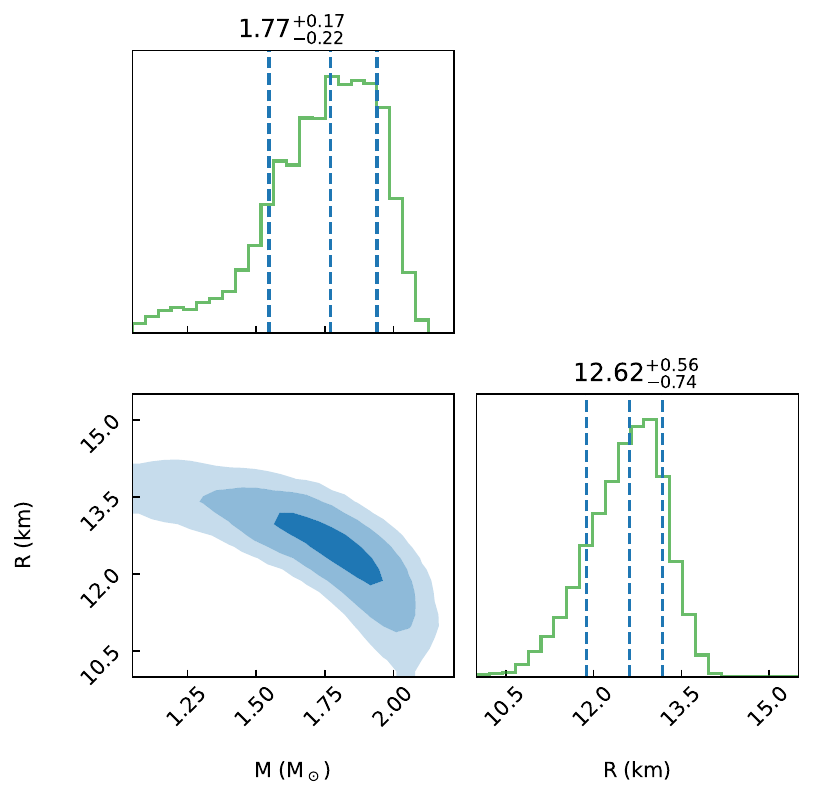}
\caption{Joint MCMC constraints on the neutron-star mass and radius of EXO~0748--676 from \textit{Chandra} and \textit{XMM-Newton} quiescent observations between 2008 and 2025, obtained with the model \texttt{TBabs*(nsatmos+powerlaw)} under the baseline distance assumption $d=7.1$~kpc. The shades from dark to light represent the $1\sigma$, $2\sigma$, and $3\sigma$ credible regions, respectively. The quoted uncertainties are $1\sigma$ credible intervals. The details of the MCMC setup and sampling procedure are described in Section~2.4. 
}
\label{fig:1}
\end{figure}

\begin{figure*}
\centering
\includegraphics[width=0.45\textwidth]{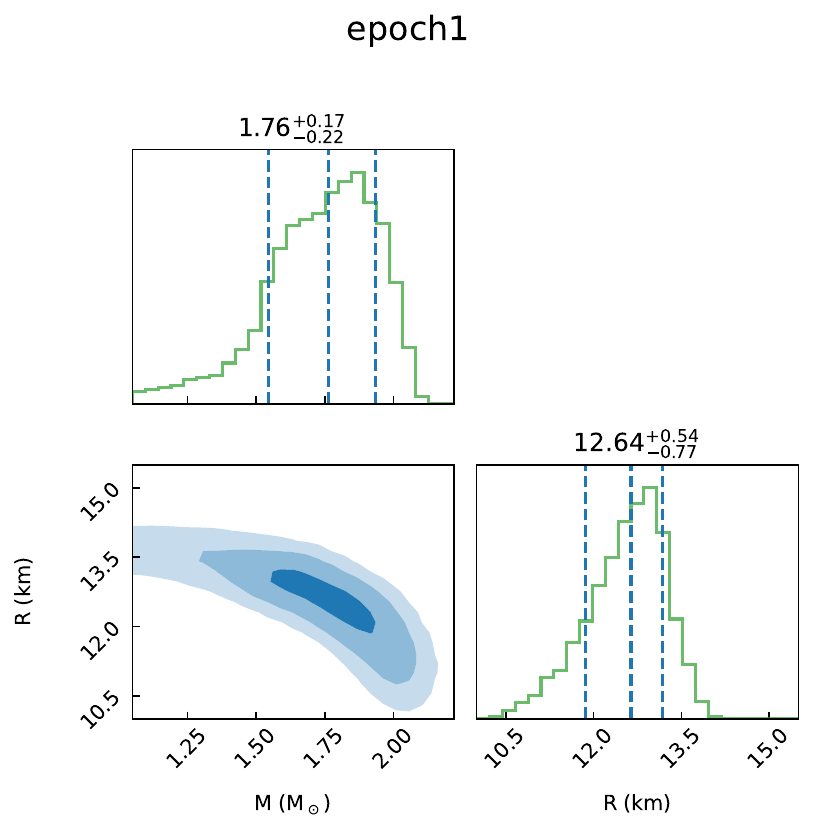}
\includegraphics[width=0.45\textwidth]{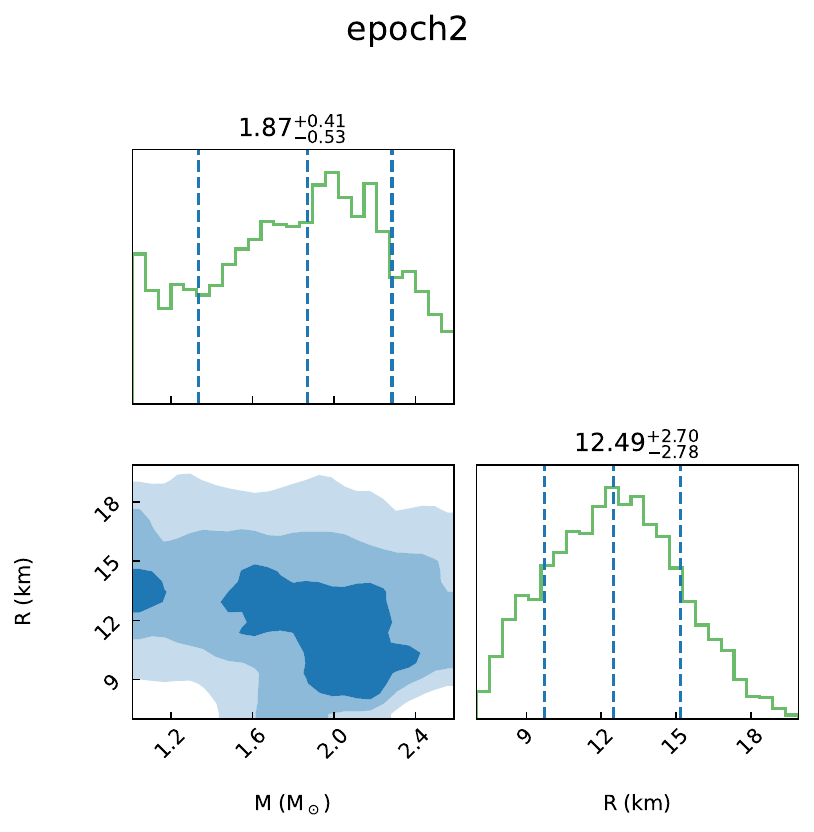}
\caption{Posterior constraints on the neutron-star mass and radius of EXO 0748$-$676 derived from observations in the first quiescent epoch (2008--2024, left panel) and the second quiescent epoch (since 2025, right panel), using the model \texttt{TBabs*(nsatmos+powerlaw)} in MCMC analyses and assuming a source distance of 7.1~kpc. The shades from dark to light represent the $1\sigma$, $2\sigma$, and $3\sigma$ credible regions, respectively. The quoted uncertainties correspond to $1\sigma$ credible intervals. Details of the MCMC configuration are provided in Section~2.3.
For the second quiescent epoch, the data do not simultaneously constrain all parameters; therefore, $N_{\rm H}$ was fixed at $0.58 \times 10^{21}\,\mathrm{cm^{-2}}$, as obtained from the fit to the full dataset.
}
\label{fig:3}
\end{figure*}

The best-fit spectral parameters, unabsorbed fluxes, and the fractions of non-thermal emission are listed in Table \ref{tab:spectral_fit_summary}. The joint analysis yields well-constrained posterior distributions for the neutron-star mass, radius, and other model parameters. In the following sections, we examine the resulting mass–radius constraints and their dependence on the adopted dataset and source distance.

\subsection{Mass-Radius Constraints}

The joint posterior distributions of the neutron-star mass and radius are shown in Figure~\ref{fig:1}. 
From the spectral fitting of quiescent data, we constrain the mass of EXO 0748$-$676 to be $1.77^{+0.17}_{-0.22}\,M_\odot$ and the radius to be $12.62^{+0.56}_{-0.74}$ km at the $1\sigma$ credible level. 
To investigate the impact of different datasets on the inferred parameters, we further performed the same joint analysis separately for the first and second quiescent epochs; the corresponding results are shown in Figure~\ref{fig:3}.

Figure \ref{fig:3} illustrates how the constraints depend on the statistical quality and constraining power of the dataset. In the first quiescent epoch, where a larger number of observations are included, the posterior distributions of both the mass and radius are relatively compact and well constrained, yielding $M = 1.76^{+0.17}_{-0.22}\,M_{\odot}$ and $R = 12.64^{+0.54}_{-0.77}$ km. 
The corresponding two-dimensional posterior shows a clear anti-correlation between mass and radius, as expected from atmosphere-model fits, but the allowed parameter space remains tightly localized around our best-fit values.

By contrast, the second quiescent epoch, which is based on a more limited dataset, exhibits substantially broader posterior distributions. In particular, the mass posterior develops a pronounced low-mass tail extending toward $\sim 1\,
M_{\odot}$, resulting in a much weaker lower bound on the neutron-star mass. This behavior indicates that the available data are insufficient to simultaneously constrain all spectral parameters with high precision. Consequently, degeneracies between the thermal atmosphere parameters, the power-law component, and the assumed hydrogen column density become significantly more important. These degeneracies broaden the allowed parameter space and reduce the constraining power of the fit.

\subsection{Non-thermal components}

The inferred power-law component is weak during both quiescent epochs and is not required in all observations. The joint fits yield photon indices of $\Gamma_1=1.86^{+0.12}_{-0.13}$ and $\Gamma_2=2.4^{+0.2}_{-0.3}$ for the first and second quiescent intervals, respectively. Several observations have power-law normalizations consistent with zero, indicating little or no detectable nonthermal emission. Given the limited number of high-energy photons and the degeneracy between the thermal and nonthermal components, the apparent difference in photon index between the two epochs should be interpreted cautiously.

\begin{table}
\centering
\caption{Results of piecewise spectral fits to the quiescent data of EXO\,0748--676, grouped by time intervals. The mass, radius, and photon index are obtained from independent fits to each epoch. Uncertainties are 1$\sigma$ credible intervals.}
\begin{tabular}{c c c c}
\toprule
MJD range & $M$ ($M_\odot$) & $R$ (km) & $\Gamma$ \\
\midrule
54751--55013 & $1.63^{+0.26}_{-0.37}$ & $12.94^{+0.69}_{-0.94}$ & $1.58^{+0.21}_{-0.26}$ \\
55306--56505 & $1.43^{+0.24}_{-0.23}$ & $12.40^{+0.38}_{-0.53}$ & $1.80^{+0.16}_{-0.18}$ \\
56397--59478 & $1.67^{+0.27}_{-0.46}$ & $12.62^{+0.76}_{-1.41}$ & $1.89^{+0.28}_{-0.37}$ \\
\bottomrule
\end{tabular}
\label{tab:piecewise}
\end{table}

Given that the non-thermal emission may vary across the data span, we divided the observations into three groups for further investigation. The first group covers MJD 54751--55013, corresponding to the early cooling phase of the first quiescent epoch (see Section~4.3). The second group spans MJD 55306--56505, covering the plateau phase and the subsequent temperature decline. The third group extends from MJD 56397 to 59478, primarily including the late-time temperature rise phase. 
Because using only the three observations with ObsIDs 0824420101, 23440, and 24516 for the third group cannot provide sufficient constraints on the photon index, we included two additional earlier observations (ObsIDs 0690330101 and 14663) in this group. We performed independent spectral fits for each group, and the resulting mass, radius, and photon index are listed in Table~\ref{tab:piecewise}. The three groups yield consistent results, indicating that the photon index did not change significantly over the entire period and supporting our approach of tying the photon index across different observations.

Another noteworthy point is that the normalization of the non-thermal radiation of this source appears to exhibit irregular variations among different observations, and the XMM-Newton normalization seems to be systematically lower than that of Chandra. It remains unclear whether these variations arise from real changes in the source or are due to insufficient signal-to-noise ratio of the data. Nevertheless, it is necessary to test their impact on the results. We tested the impact of typical variations in the non-thermal radiation on the fitting results by fixing the powerlaw normalization at 0 and at \(1\times10^{-5}\,\mathrm{ph\,keV^{-1}\,cm^{-2}\,s^{-1}}\), respectively. The latter value is a representative order of magnitude for the non-thermal component in this source and comparable to the values inferred from Chandra observations. While fixing all other parameters at their best-fit values and fitting only the temperature, the test was performed on observation 0605560401, because it yields a non-thermal fraction of \(1\%\), which is significantly smaller than that of other contemporaneous observations. The results show that this causes a temperature difference of about \(0.0011\,\mathrm{keV}\), which is acceptable given our fitting precision. The same method was applied to the mass and radius, yielding a mass difference of \(0.08\,M_\odot\) and a radius difference of \(0.17\,\mathrm{km}\), respectively. Compared with the temperature, the impact on the mass and radius is somewhat larger, reaching a few percent (1-5\%), but it is within an acceptable range under the current constraint precision. At the same time, joint fitting also weakens the impact of uncertainties from individual spectra.

\subsection{Systematic Uncertainties Due to Distance Uncertainty}

\begin{figure}
\centering
\includegraphics[width=0.45\textwidth]{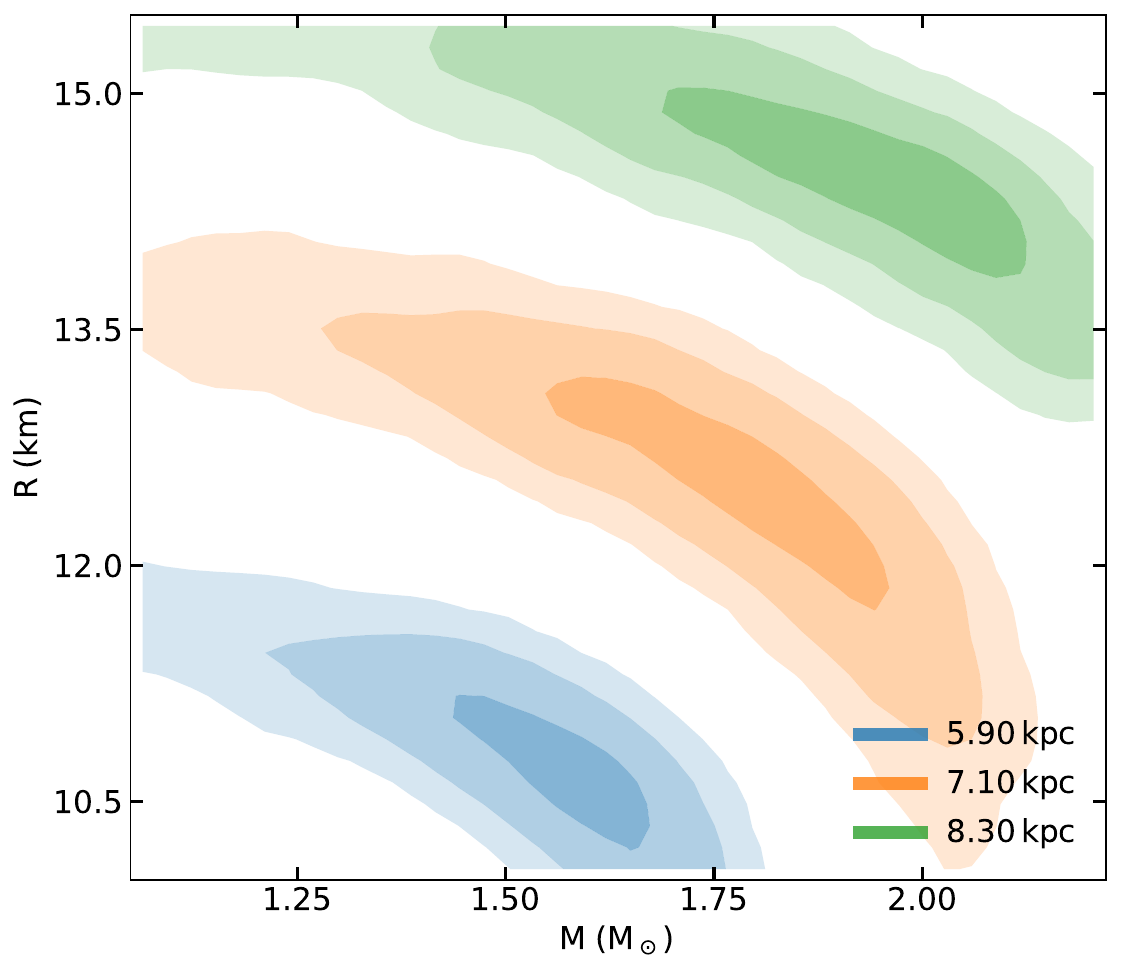}
\caption{Posterior distributions for $M,R$ as the distance is fixed at $d=5.90,~7.10,~8.30\,\mathrm{kpc}$.
  Different colors correspond to different distances,
  and within each color the shades from dark to light represent
  the $1\sigma$, $2\sigma$, and $3\sigma$ contours, respectively.
}
\label{fig:mr_vs_d_contour}
\end{figure}

The dominant systematic uncertainty in atmosphere-based mass–radius inference is the source distance.
\cite{2008MNRAS.387..268G} reported a distance of $7.1\pm1.2\,\mathrm{kpc}$. 
To evaluate the effect of this uncertainty on the inferred neutron-star parameters, we performed spectral fits over a broad distance range of $5.9$--$8.3$~kpc. The resulting mass and radius posterior distributions obtained under different distance assumptions are shown in Figure \ref{fig:mr_vs_d_contour}. 
We find that both $M$ and $R$ increase monotonically with increasing distance. At $d = 5.9$ kpc, we obtain $M = 1.57^{+0.12}_{-0.16}\,M_\odot$ and $R = 10.73^{+0.48}_{-0.58}$ km, while at $d = 8.3$ kpc the inferred values become $M = 1.91^{+0.20}_{-0.28}\,M_\odot$ and $R = 14.57^{+0.56}_{-0.63}$ km.
This trend is expected because a larger source distance implies a higher intrinsic luminosity for the observed thermal flux, which in turn requires a larger emitting area and stronger gravitational effects in the atmosphere-model fits.
When accounting for the full allowed distance range, we therefore obtain conservative constraints of  $M\simeq 1.41-2.11 M_{\odot}$ and $R\simeq 10.15-15.13$ km, which we adopt as our final systematic uncertainties.

\begin{table*}[ht]
\centering
\caption{
Constraints on the neutron-star mass and radius of EXO 0748$-$676 obtained using different methods. 
}
\renewcommand{\arraystretch}{1.1}
\setlength{\tabcolsep}{3pt}
\begin{tabular}{lccc}
\hline\hline
Method & $M~(M_{\odot})$ & $R~(\mathrm{km})$ & Reference \\
\hline
Gravitational redshift ($z=0.35$) 
& $1.4$--$1.8$ 
& $9$--$12$ 
& \cite{2002Natur.420...51C} \\
Eddington limit + redshift + cooling tail
& $2.10 \pm 0.28$ 
& $13.8 \pm 1.8$ 
& \cite{2006Natur.441.1115O} \\
Dynamical constraints 
& $>1.5$ 
& --- 
& \cite{2009MNRAS.394L.136M} \\
\textit{XMM} spectral fitting (\texttt{nsagrav} / \texttt{nsatmos}) 
& $1.77 \pm 0.45 / 1.71 \pm 0.30$ 
& $16.6^{+1.8}_{-7.5} / 16.5 \pm 0.5$ 
& \cite{2011MNRAS.414.1077Z} \\
\textit{XMM} spectral fitting 
& $1.78^{+0.4}_{-0.6}$ 
& $13.7^{+1.0}_{-2.7}$ 
& \cite{2011AA...528A.150D} \\
Optical spectroscopy 
& $>1.27$ 
& --- 
& \cite{2012MNRAS.420...75R} \\
\textit{Chandra} spectral fitting 
& $1.64 \pm 0.38$ 
& $13.2^{+0.6}_{-2.0}$ 
& \cite{2014ApJ...791...47D} \\
\textit{XMM} spectral fitting (0.3--10 / 0.5--10 keV) 
& $2.08^{+0.07}_{-0.15} / 1.50^{+0.40}_{-0.99}$ 
& $11.9 \pm 0.7 / 12.2^{+1.0}_{-3.6}$ 
& \cite{2017MNRAS.471.2605C} \\
Eclipse mapping (Gaussian/Exponential density profile)
& $2.01^{+0.22}_{-0.21} / 2.02^{+0.29}_{-0.27}$ 
& --- 
& \cite{2022MNRAS.510.4736K} \\
Quiescent spectral fitting (\textit{Chandra+XMM}; \texttt{nsatmos})
& $1.77^{+0.17}_{-0.22}$ 
& $12.62^{+0.56}_{-0.74}$ 
& This work (distance $\simeq 7.1$ kpc)\\
\hline
\end{tabular}
\end{table*}

\section{Discussion}
\label{sec:discussion}

Our analysis of the complete quiescent dataset for EXO 0748$-$676 yields improved constraints on the neutron-star mass and radius. Although systematic uncertainties, particularly those associated with the source distance, remain important, the inferred mass–radius region is broadly consistent with relatively stiff dense-matter EOSs. We now place these results in the context of previous measurements and current constraints on neutron-star structure.
We also discuss the temperature evolution of this source in light of the inferred mass.

\subsection{Synthesis and Comparison of Results}
\label{sec:synthesis}

Table~3 summarizes the constraints on the mass and radius of EXO~0748$-$676 obtained with a variety of observational techniques, including gravitational-redshift measurements, burst spectroscopy, optical dynamical studies, and previous quiescent spectral analyses. 
Our joint analysis of the combined \textit{Chandra} and \textit{XMM-Newton}  quiescent dataset yields results that are broadly consistent with earlier studies but substantially reduce the allowed parameter space.

Dynamical constraints from optical spectroscopy provide an independent lower bound on the mass. \citet{2009MNRAS.394L.136M} used Doppler tomography of narrow emission components in the He II lines to measure the companion star's center-of-mass radial-velocity semi-amplitude $K_2 > 300$~km~s$^{-1}$ 
and obtain $1.0 < M/M_{\odot} < 2.4$, with $M > 1.5\,M_{\odot}$ if the donor is on the main sequence.
Our posterior mass distribution is fully consistent with this range and supports the conclusion that the neutron star in EXO 0748$-$676 is relatively massive. Eclipse mapping combined with radial-velocity constraints by
\citet{2022MNRAS.510.4736K} suggests $M = 2.01^{+0.22}_{-0.21}\,M_{\odot}$ and $2.02^{+0.29}_{-0.27}\,M_{\odot}$ assuming Gaussian and exponential density profiles for the absorbing material, respectively, which is somewhat higher than our best-fit value but compatible within the combined uncertainties.

Thermal spectral fits to quiescent \textit{XMM-Newton} and \textit{Chandra} data have previously yielded a range of mass-radius estimates for EXO 0748$-$676. 
The combination of thermal spectral fits and distance estimates allows constraints on the neutron-star radius. \citet{2011MNRAS.414.1077Z} used \texttt{nsagrav} and \texttt{nsatmos} models to infer $M \approx 1.7-1.8\,M_{\odot}$ and radii $R \approx 13-17$ km, with large systematic uncertainties dominated by the distance. 
\citet{2014ApJ...791...47D} found $M = 1.64 \pm 0.38\,M_{\odot}$ and $R = 13.2^{+0.6}_{-2.0}$ km from \textit{Chandra} spectra, 
while \citet{2011AA...528A.150D} obtained $M = 1.78^{+0.4}_{-0.6}\,M_{\odot}$ and $R = 13.7^{+1.0}_{-2.7}$ km.
\citet{2017MNRAS.471.2605C} reported somewhat model- and band-dependent results, with radii in the range $R \sim 12-14$ km. As summarized in Table 3, all of these works favor relatively large radii, and our combined \textit{Chandra+XMM} constraints refine this picture by significantly reducing the allowed parameter space: our $1\sigma$ region overlaps the higher-radius portions of earlier results and excludes very small radii.

Additional constraints have been derived from thermonuclear bursts and candidate gravitationally redshifted absorption lines. \citet{2002Natur.420...51C} reported a gravitational redshift $z = 0.35$ from narrow absorption features in \textit{XMM-Newton} RGS spectra, implying $M \sim 1.4-1.8\,M_{\odot}$ and $R \sim 9-12 $ km. \citet{2006Natur.441.1115O} combined this redshift measurement with the Eddington limit and cooling-tail modeling of Type I X-ray bursts to infer $M = 2.10 \pm 0.28\,M_{\odot}$ and $R = 13.8 \pm 1.8$ km. 
However, \citet{lin2010} showed that the narrow line widths reported by \citet{2002Natur.420...51C} are difficult to reconcile with the 552 Hz spin frequency discovered in the rising phase of two Type-I bursts \citep{2010ApJ...711L.148G} and with the high burst oscillation amplitudes, casting doubt on a photospheric origin for the lines and thus on the robustness of the corresponding redshift constraint.

Overall, the combined \textit{Chandra} and \textit{XMM-Newton} quiescent spectra yield mass and radius estimates that are consistent with independent dynamical and burst-based constraints, while providing substantially tighter statistical uncertainties than most previous X-ray spectral analyses. The relatively high mass and radius inferred here favor a stiff dense-matter EOS and disfavor very compact configurations with $R \lesssim 10$ km. These implications are discussed further in Section 4.2.

\subsection{Equation of State Implications}

The structure of a non-rotating neutron star in hydrostatic equilibrium is determined by solving the Tolman--Oppenheimer--Volkoff (TOV) equations, given an EOS that specifies the pressure as a function of density. The EOS at supranuclear densities is poorly constrained from first principles because it depends on the detailed behavior of the strong nuclear force in a many-body environment~\citep[e.g.,][]{2013ApJ...765L...5S,2022NatRP...4..237Y}.
Different theoretical approaches---ranging from non-relativistic potential models to relativistic mean-field theories and chiral effective field theory calculations---predict a range of possible EOSs, each corresponding to a distinct $M$--$R$ curve \citep[e.g.,][]{2014ApJ...784..123L, 2019ApJ...887...48B}.

The $M$--$R$ relation encodes crucial information about the stiffness of the EOS. A stiff EOS, characterized by a rapid rise in pressure with density, produces neutron stars with larger radii for a given mass. Conversely, a soft EOS yields more compact stars with smaller radii. Exotic phases of matter---such as hyperons, Bose--Einstein condensates of pions or kaons, or deconfined quark matter---tend to soften the EOS, reducing the maximum mass and radius \citep[e.g.,][]{2020JHEAp..28...19L}. Therefore, measuring the $M$--$R$ relation observationally provides a direct test of whether such exotic states exist in neutron-star cores.

Our mass-radius constraints for EXO 0748$-$676 contribute an important data point to this effort. 
Previous atmosphere-model fits to quiescent \textit{XMM-Newton} data for this source led \citet{2011MNRAS.414.1077Z} to exclude the soft quark-matter EOS SQM1 for the known distance range. More broadly, studies of quiescent LMXBs have shown that EOSs predicting very small radii $R \lesssim 10$ km are disfavored  \citep{2014ApJ...796L...3G, 2019ApJ...887...48B}. 
Our combined \textit{Chandra+XMM} analysis strengthens this picture: across the full distance range $d = 5.9$–8.3~kpc explored in Section~3.3, the inferred radii remain $R \gtrsim 10.15$~km even at the smallest distance, thereby disfavoring very soft EOSs that predict neutron stars with $R \lesssim 10$~km at $M \sim 1.41$–2.11~M$_\odot$.
Independent dynamical constraints also point toward a relatively high neutron-star mass in EXO 0748$-$676~\citep{2009MNRAS.394L.136M,2022MNRAS.510.4736K}. These estimates are consistent with the relatively high mass inferred from our quiescent spectral fits and with the requirement that the EOS support $\sim 2\,M_{\odot}$ neutron stars.

Our results also fit within the broader landscape of recent constraints from pulse profile modeling with the Neutron Star Interior Composition Explorer (NICER) and from gravitational-wave observations. NICER measurements of PSR J0030+0451 and PSR J0740+6620 indicate radii $R \sim 12-14$ km for neutron stars with masses $\sim 1.4-2.1 M_{\odot}$  \citep{2019ApJ...887L..21R, 2019ApJ...887L..24M, 2021ApJ...918L..28M, 2021ApJ...918L..27R}, while gravitational-wave tidal deformability measurements favor EOSs that are neither extremely soft nor extremely stiff \citep{2022NatRP...4..237Y}. 
From \textit{NICER+XMM}+GW170817 posteriors, \citet{2024PhRvD.109l3005M} has reported $R_{1.4}=12.82_{-0.46}^{+0.36}$ km for typical $1.4M_{\odot}$ neutron stars, and $12.85_{-0.51}^{+0.45}$ km when allowing a transition to deconfined quark phase in their interior. 
To constrain fine features of the EOS (e.g., the presence and onset density of phase transitions, or nontrivial behavior of the speed of sound), it is necessary to combine thermal X‑ray results with pulse-profile modeling, burst cooling/burst-tail methods, dynamical/doppler constraints, and gravitational-wave tidal deformability measurements in a self-consistent Bayesian framework. \citet{2024PhRvD.109l3005M} showed that such joint inference changes the posterior ranges for microphysical parameters only modestly in current data, but the constraining power will improve with higher-quality, cross-calibrated observations.

\subsection{Temperature evolution}

\begin{figure}
\centering
\includegraphics[width=0.45\textwidth]{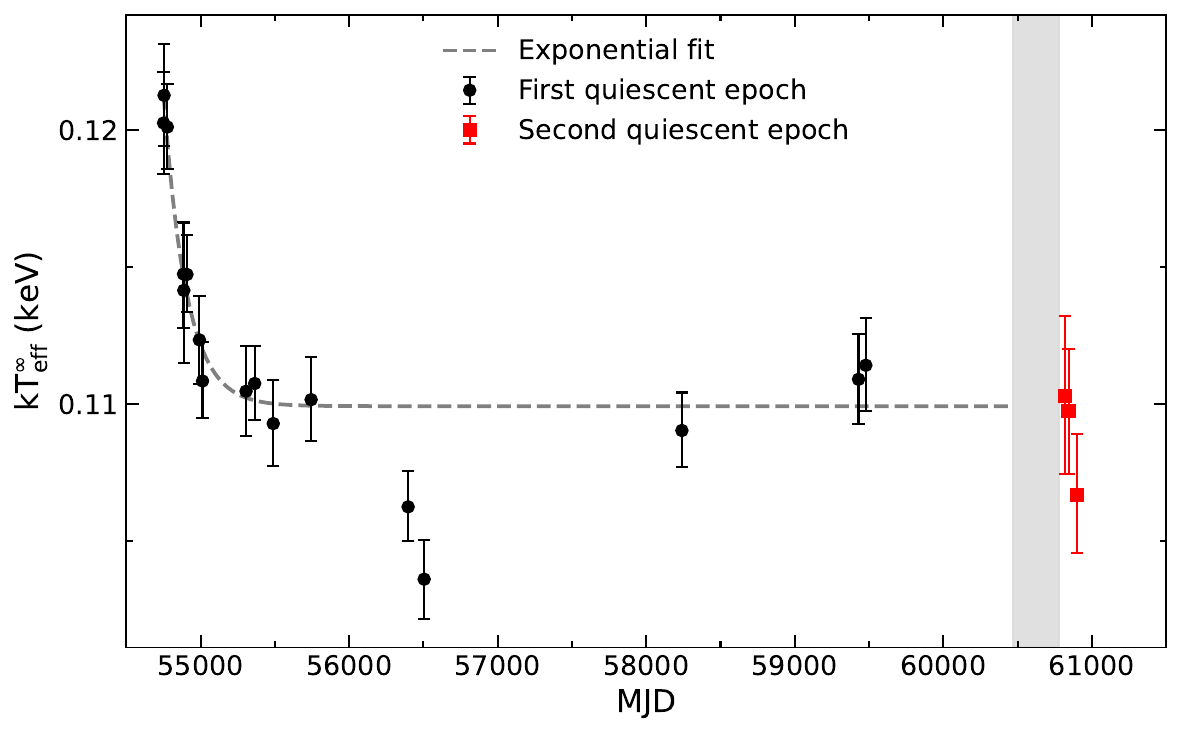}
\caption{Evolution of the effective temperature in the two quiescent epochs of EXO 0748$-$676.
Black points show quiescent observations between 2008 and 2021, following the 1985--2008 outburst (first quiescent epoch). Red points show early quiescent observations in 2025, following the 2024--2025 outburst (second quiescent epoch).
The shaded region marks the 2024--2025 outburst.
Error bars indicate 1$\sigma$ uncertainties obtained under the inferred neutron-star mass of $M = 1.77\,M_\odot$.
}
\label{fig:Teff}
\end{figure}

The quiescent spectra of EXO 0748$-$676 are dominated by thermal emission from the neutron-star surface, allowing the long-term evolution of the effective temperature to be tracked over nearly two decades.
To study this evolution, we fixed the neutron-star mass $M$ in the \texttt{nsatmos} model at the best-fit value obtained above, $1.77\,M_\odot$, while keeping all other parameter settings identical to those described in Section~2.3, and refit the spectra. The resulting evolution of the effective temperature is shown in Figure~\ref{fig:Teff}. We have converted the temperatures to the redshifted temperatures according to \(kT_{\rm eff}^{\infty}=kT_{\rm eff}/(1+z)\), where \(1+z=(1-R_s/R)^{-1/2}\) and \(R_s=2GM/c^2\) is the Schwarzschild radius.

In the first quiescent epoch, following the end of the 2008 outburst, the effective temperature declined rapidly from $\sim0.121$\,keV to $\sim0.110$\,keV within the first $\sim260$ days, consistent with thermal relaxation of an accretion-heated crust. 
Thereafter, the temperature remained approximately constant at $\simeq0.110$\,keV for at least $\sim700$ days. 
This plateau phase has been interpreted as the result of compositionally driven convection associated with chemical separation during crust crystallization, which transports heat inward and temporarily delays the cooling of the outer layers~\citep{2017MNRAS.471.2605C}.

Later, in two observations taken in April and August 2013, the effective temperature of the source showed a sharp drop. 
A particularly intriguing feature of the cooling curve is the late-time temperature increase reported by \citet{2020A&A...638L...2P} and supported by subsequent \textit{Chandra} observations. Such behavior is not predicted by standard crust-cooling models and is difficult to explain solely through low-level residual accretion, leaving the physical origin of the reheating episode uncertain. This anomalous evolution suggests that additional heating mechanisms or more complex crustal physics may be operating in EXO 0748$-$676. One geometric effect worth considering is that whether the emission originates from the entire neutron-star surface also affects the inferred temperature. From a purely observational perspective, if the emission arises only from a fraction of the surface during the early cooling phase and later covers the entire surface, the early temperature would be systematically underestimated, making the later temperature appear anomalously elevated. However, the current observational precision is insufficient to distinguish whether the emission comes from the whole surface, and there is currently no clear theoretical mechanism that would cause the effective emitting area to evolve over time.

We also fitted the early cooling curve of the first quiescent epoch with a simple exponential decay model, \(kT(t) = A \exp[-(t - t_0)/\tau] + C\). Given the anomalous evolution of the source during the late quiescent epoch, we excluded the observations showing the anomalous temperature rise (ObsID 0824420101, 23440, 24516) and the two slightly earlier temperature drop observations (ObsID 0690330101, 14663) from the fit, in order to better fit the early cooling curve of the quiescent epoch. The fit yielded \(\tau = 153.6 \pm 43.9\) days, \(C = 0.10993 \pm 0.00086\) keV, and \(A = 0.0144 \pm 0.0019\) keV, which is broadly consistent with the earlier fitting results of \citep{2011MNRAS.412.1409D}. We then extrapolated the best-fit curve and found that the temperature in the second quiescent epoch does not show a significant increase compared to the late stage of the first quiescent epoch. This indicates that the crustal heating during the second accretion phase was weaker, far less than that during the first phase.

The recent return of the source to quiescence following the 2024–2025 outburst provides a valuable opportunity to test these ideas. Early \textit{Chandra} observations already suggest a possible renewed decline in the effective temperature, consistent with the onset of a new phase of crustal cooling. Continued monitoring will reveal whether the thermal evolution follows the pattern observed after the 2008 outburst, including the appearance of a plateau or a late-time temperature increase. Such observations will provide important constraints on the thermal properties, composition, and heat transport processes in the neutron-star crust.

\section{Summary andy Outlook}
\label{sec:summary}

The neutron star EXO~0748$-$676 has long served as a valuable laboratory for probing the neutron-star mass-radius relation and the dense-matter EOS. 
We have performed a joint analysis of all available quiescent \textit{Chandra} and \textit{XMM-Newton} observations of EXO~0748$-$676 obtained between 2008 and 2025, spanning two quiescent epochs separated by nearly 16 years. 
Our new constraints on its mass and radius reinforce this role, pointing toward relatively stiff EOSs that produce radii $ \gtrsim 10$ km for neutron stars in the $\sim 1.41-2.11\,M_{\odot} $ range.

At the same time, systematic model uncertainties remain significant. Variations in atmosphere composition (hydrogen versus helium or metals), magnetic field strength, assumptions about surface temperature uniformity, cross-calibration between instruments, and the adopted source distance can each shift inferred radii by $\sim0.5-1.5$ km in individual studies. In this work we have obtained tighter constraints by adopting (i) a physically motivated hydrogen-atmosphere model and (ii) a joint fitting approach that ties global parameters across observations. We note that these constraints are conditional on these assumptions, and the associated systematic uncertainties must be explicitly incorporated in EOS-level inference. 

Also, the recent return of EXO 0748$-$676 to outburst provides a valuable opportunity to investigate a new episode of crust cooling and to compare the thermal evolution following two long-duration accretion events.
The effective temperature evolution over nearly two decades reveals complex behavior, including an initial rapid cooling phase~\citep{2011MNRAS.412.1409D}, a plateau~\citep{2014ApJ...791...47D}, late-time reheating before the 2024--2025 outburst~\citep{2020A&A...638L...2P}, and a declining temperature trend that appears to be indicated by the early post-outburst data. 
This suggests that additional heating mechanisms or more complex crustal physics may operate in EXO~0748$-$676 beyond standard crust-cooling scenarios~\citep{2014ApJ...791...47D,2017MNRAS.471.2605C,2020A&A...638L...2P}.
Continued monitoring of the source in quiescence will help constrain the thermal properties of the neutron-star crust, while deeper observations will improve measurements of the weak nonthermal component and further refine atmosphere-model constraints.

\section*{Acknowledgments}
The work is supported by the National Natural Science Foundation of China (grant Nos. 12494572, 12273028).
The work is based on observations obtained with \textit{Chandra} X-ray Observatory available through the Chandra Data Collection (DOI: 10.25574/cdc.629). It also makes use of observations obtained with \textit{XMM-Newton}, an ESA science mission with instruments and contributions directly funded by ESA member states and NASA. 

\bibliographystyle{aasjournal}
\bibliography{references}

@ARTICLE{2020JHEAp..28...19L,
       author = {{Li}, A. and {Zhu}, Z.-Y. and {Zhou}, E.-P. and {Dong}, J.-M. and {Hu}, J.-N. and {Xia}, C.-J.},
        title = "{Neutron star equation of state: Quark mean-field (QMF) modeling and applications}",
      journal = {Journal of High Energy Astrophysics},
         year = 2020,
        month = nov,
       volume = {28},
        pages = {19-46},
          doi = {10.1016/j.jheap.2020.07.001},
archivePrefix = {arXiv},
       eprint = {2007.05116},
 primaryClass = {nucl-th},
       adsurl = {https://ui.adsabs.harvard.edu/abs/2020JHEAp..28...19L}
}

@ARTICLE{2025SCPMA..6819503L,
       author = {{Li}, Ang and {Watts}, Anna L. and {Zhang}, Guobao and {Guillot}, Sebastien and {Xu}, Yanjun and {Santangelo}, Andrea and {Zane}, Silvia and {Feng}, Hua and {Zhang}, Shuang-Nan and {Ge}, Mingyu and {Qi}, Liqiang and {Salmi}, Tuomo and {Dorsman}, Bas and {Miao}, Zhiqiang and {Tu}, Zhonghao and {Cavecchi}, Yuri and {Zhou}, Xia and {Zheng}, Xiaoping and {Wang}, Weihua and {Cheng}, Quan and {Liu}, Xuezhi and {Wei}, Yining and {Wang}, Wei and {Xu}, Yujing and {Weng}, Shanshan and {Zhu}, Weiwei and {Li}, Zhaosheng and {Shao}, Lijing and {Tuo}, Youli and {Dohi}, Akira and {Lyu}, Ming and {Liu}, Peng and {Yuan}, Jianping and {Wang}, Mingyang and {Zhang}, Wenda and {Li}, Zexi and {Tao}, Lian and {Zhang}, Liang and {Shen}, Hong and {Provid{\^e}ncia}, Constan{\c{c}}a and {Tolos}, Laura and {Patruno}, Alessandro and {Li}, Li and {Liu}, Guozhu and {Zhou}, Kai and {Chen}, Lie-Wen and {Fan}, Yizhong and {Kajino}, Toshitaka and {Lai}, Dong and {Li}, Xiangdong and {Meng}, Jie and {Tang}, Xiaodong and {Xiao}, Zhigang and {Xiong}, Shaolin and {Xu}, Renxin and {Zhou}, Shan-Gui and {Ballantyne}, David R. and {Burgio}, G. Fiorella and {Chenevez}, J{\'e}r{\^o}me and {Choudhury}, Devarshi and {Fantina}, Anthea F. and {Galloway}, Duncan K. and {Gulminelli}, Francesca and {Hebeler}, Kai and {Hoogkamer}, Mariska and {Horvath}, Jorge E. and {Kini}, Yves and {Kurkela}, Aleksi and {Linares}, Manuel and {Margueron}, J{\'e}r{\^o}me and {Mendes}, Melissa and {Oertel}, Micaela and {Papitto}, Alessandro and {Poutanen}, Juri and {Rea}, Nanda and {Schwenk}, Achim and {Song}, Xin-Ying and {Svensson}, Isak and {Tsang}, David and {Vuorinen}, Aleksi and {Andersson}, Nils and {Miller}, M. Coleman and {Rezzolla}, Luciano and {Stone}, Jirina R. and {Thomas}, Anthony W.},
        title = "{Dense matter in neutron stars with eXTP}",
      journal = {Science China Physics, Mechanics, and Astronomy},
         year = 2025,
        month = sep,
       volume = {68},
       number = {11},
          eid = {119503},
        pages = {119503},
          doi = {10.1007/s11433-025-2761-4},
archivePrefix = {arXiv},
       eprint = {2506.08104},
 primaryClass = {astro-ph.HE},
       adsurl = {https://ui.adsabs.harvard.edu/abs/2025SCPMA..6819503L}
}

@article{parikh2021,
  author = {Parikh, A. S. and Degenaar, N.},
  title = {{UV} and {X}-ray observations of the neutron star {LMXB} {EXO} 0748-676 in its quiescent state},
  journal = {Monthly Notices of the Royal Astronomical Society},
  volume = {501},
  pages = {1453--1464},
  year = {2021},
  doi = {10.1093/mnras/staa3734}
}

@ARTICLE{2024arXiv241006201S,
       author = {{Subba}, Nirpat and {Subba}, Nishika and {Paul}, Jyoti and {Sharma}, Pankaj and {Ghimiray}, Monika},
        title = "{Eclipse Dynamics and X-ray Burst Characteristics in the Low-Mass X-ray Binary EXO 0748-676}",
      journal = {arXiv e-prints},
         year = 2024,
        month = oct,
          eid = {arXiv:2410.06201},
        pages = {arXiv:2410.06201},
          doi = {10.48550/arXiv.2410.06201},
archivePrefix = {arXiv},
       eprint = {2410.06201},
 primaryClass = {astro-ph.HE},
       adsurl = {https://ui.adsabs.harvard.edu/abs/2024arXiv241006201S}
}

@article{lin2010,
  author = {Lin, J. and {\"O}zel, F. and Chakrabarty, D. and Psaltis, D.},
  title = {The incompatibility of rapid rotation with narrow photospheric {X}-ray lines in {EXO}~0748--676},
  journal = {The Astrophysical Journal},
  volume = {723},
  number = {2},
  pages = {1053--1059},
  year = {2010},
  doi = {10.1088/0004-637X/723/2/1053}
}

@ARTICLE{1999AJ....118.1390G,
       author = {{Garcia}, Michael R. and {Callanan}, Paul J.},
        title = "{Two Neutron Star Soft X-Ray Transients in Quiescence: 4U 2129+47 and EXO 0748-676}",
      journal = {\aj},
         year = 1999,
        month = sep,
       volume = {118},
       number = {3},
        pages = {1390-1394},
          doi = {10.1086/301014},
archivePrefix = {arXiv},
       eprint = {astro-ph/9904117},
 primaryClass = {astro-ph},
       adsurl = {https://ui.adsabs.harvard.edu/abs/1999AJ....118.1390G}
}

@ARTICLE{2014ApJ...784..123L,
       author = {{Lattimer}, James M. and {Steiner}, Andrew W.},
        title = "{Neutron Star Masses and Radii from Quiescent Low-mass X-Ray Binaries}",
      journal = {\apj},
         year = 2014,
        month = apr,
       volume = {784},
       number = {2},
          eid = {123},
        pages = {123},
          doi = {10.1088/0004-637X/784/2/123},
archivePrefix = {arXiv},
       eprint = {1305.3242},
 primaryClass = {astro-ph.HE},
       adsurl = {https://ui.adsabs.harvard.edu/abs/2014ApJ...784..123L}
}

@ARTICLE{2014ApJ...796L...3G,
       author = {{Guillot}, Sebastien and {Rutledge}, Robert E.},
        title = "{Rejecting Proposed Dense Matter Equations of State with Quiescent Low-mass X-Ray Binaries}",
      journal = {\apjl},
         year = 2014,
        month = nov,
       volume = {796},
       number = {1},
          eid = {L3},
        pages = {L3},
          doi = {10.1088/2041-8205/796/1/L3},
archivePrefix = {arXiv},
       eprint = {1409.4306},
 primaryClass = {astro-ph.HE},
       adsurl = {https://ui.adsabs.harvard.edu/abs/2014ApJ...796L...3G}
}

@ARTICLE{2016RvMP...88b1001W,
       author = {{Watts}, Anna L. and {Andersson}, Nils and {Chakrabarty}, Deepto and {Feroci}, Marco and {Hebeler}, Kai and {Israel}, Gianluca and {Lamb}, Frederick K. and {Miller}, M. Coleman and {Morsink}, Sharon and {{\"O}zel}, Feryal and {Patruno}, Alessandro and {Poutanen}, Juri and {Psaltis}, Dimitrios and {Schwenk}, Achim and {Steiner}, Andrew W. and {Stella}, Luigi and {Tolos}, Laura and {van der Klis}, Michiel},
        title = "{Colloquium: Measuring the neutron star equation of state using x-ray timing}",
      journal = {Reviews of Modern Physics},
         year = 2016,
        month = apr,
       volume = {88},
       number = {2},
          eid = {021001},
        pages = {021001},
          doi = {10.1103/RevModPhys.88.021001},
archivePrefix = {arXiv},
       eprint = {1602.01081},
 primaryClass = {astro-ph.HE},
       adsurl = {https://ui.adsabs.harvard.edu/abs/2016RvMP...88b1001W}
}

@ARTICLE{2013ApJ...765L...5S,
       author = {{Steiner}, Andrew W. and {Lattimer}, James M. and {Brown}, Edward F.},
        title = "{The Neutron Star Mass-Radius Relation and the Equation of State of Dense Matter}",
      journal = {\apjl},
         year = 2013,
        month = mar,
       volume = {765},
       number = {1},
          eid = {L5},
        pages = {L5},
          doi = {10.1088/2041-8205/765/1/L5},
archivePrefix = {arXiv},
       eprint = {1205.6871},
 primaryClass = {nucl-th},
       adsurl = {https://ui.adsabs.harvard.edu/abs/2013ApJ...765L...5S}
}

@ARTICLE{2009A&A...500..891S,
       author = {{Suleimanov}, V. and {Potekhin}, A.~Y. and {Werner}, K.},
        title = "{Models of magnetized neutron star atmospheres: thin atmospheres and partially ionized hydrogen atmospheres with vacuum polarization}",
      journal = {\aap},
         year = 2009,
        month = jun,
       volume = {500},
       number = {2},
        pages = {891-899},
          doi = {10.1051/0004-6361/200912121},
archivePrefix = {arXiv},
       eprint = {0905.3276},
 primaryClass = {astro-ph.SR},
       adsurl = {https://ui.adsabs.harvard.edu/abs/2009A&A...500..891S}
}

@INPROCEEDINGS{2002nsps.conf..263Z,
       author = {{Zavlin}, V.~E. and {Pavlov}, G.~G.},
        title = "{Modeling Neutron Star Atmospheres}",
    booktitle = {Neutron Stars, Pulsars, and Supernova Remnants},
         year = 2002,
       editor = {{Becker}, W. and {Lesch}, H. and {Tr{\"u}mper}, J.},
        month = jan,
        pages = {263},
          doi = {10.48550/arXiv.astro-ph/0206025},
archivePrefix = {arXiv},
       eprint = {astro-ph/0206025},
 primaryClass = {astro-ph},
       adsurl = {https://ui.adsabs.harvard.edu/abs/2002nsps.conf..263Z}
}

@ARTICLE{1998A&ARv...8..279C,
       author = {{Campana}, S. and {Colpi}, M. and {Mereghetti}, S. and {Stella}, L. and {Tavani}, M.},
        title = "{The neutron stars of Soft X-ray Transients}",
      journal = {\aapr},
         year = 1998,
        month = jan,
       volume = {8},
       number = {4},
        pages = {279-316},
          doi = {10.1007/s001590050012},
archivePrefix = {arXiv},
       eprint = {astro-ph/9805079},
 primaryClass = {astro-ph},
       adsurl = {https://ui.adsabs.harvard.edu/abs/1998A&ARv...8..279C}
}

@ARTICLE{2022NatRP...4..237Y,
       author = {{Yunes}, Nicol{\'a}s and {Miller}, M. Coleman and {Yagi}, Kent},
        title = "{Gravitational-wave and X-ray probes of the neutron star equation of state}",
      journal = {Nature Reviews Physics},
         year = 2022,
        month = feb,
       volume = {4},
       number = {4},
        pages = {237-246},
          doi = {10.1038/s42254-022-00420-y},
archivePrefix = {arXiv},
       eprint = {2202.04117},
 primaryClass = {gr-qc},
       adsurl = {https://ui.adsabs.harvard.edu/abs/2022NatRP...4..237Y}
}

@ARTICLE{2019ApJ...887...48B,
       author = {{Baillot d'Etivaux}, Nicolas and {Guillot}, Sebastien and {Margueron}, J{\'e}r{\^o}me and {Webb}, Natalie and {Catelan}, M{\'a}rcio and {Reisenegger}, Andreas},
        title = "{New Constraints on the Nuclear Equation of State from the Thermal Emission of Neutron Stars in Quiescent Low-mass X-Ray Binaries}",
      journal = {\apj},
         year = 2019,
        month = dec,
       volume = {887},
       number = {1},
          eid = {48},
        pages = {48},
          doi = {10.3847/1538-4357/ab4f6c},
archivePrefix = {arXiv},
       eprint = {1905.01081},
 primaryClass = {astro-ph.HE},
       adsurl = {https://ui.adsabs.harvard.edu/abs/2019ApJ...887...48B}
}

@INPROCEEDINGS{2008AIPC..983..519J,
       author = {{Jonker}, P.~G.},
        title = "{Constraining the neutron star equation of state using quiescent low-mass X-ray binaries}",
    booktitle = {40 Years of Pulsars: Millisecond Pulsars, Magnetars and More},
         year = 2008,
       editor = {{Bassa}, C. and {Wang}, Z. and {Cumming}, A. and {Kaspi}, V.~M.},
       series = {American Institute of Physics Conference Series},
       volume = {983},
        month = feb,
    publisher = {AIP},
        pages = {519-525},
          doi = {10.1063/1.2900287},
archivePrefix = {arXiv},
       eprint = {0711.2579},
 primaryClass = {astro-ph},
       adsurl = {https://ui.adsabs.harvard.edu/abs/2008AIPC..983..519J}
}

@MISC{2004cxo..prop.1673W,
       author = {{Wijnands}, Rudy},
        title = "{Cooling curves of accretion-heated neutron stars}",
 howpublished = {Chandra Proposal ID 06400229},
         year = 2004,
        month = sep,
        pages = {1673},
       adsurl = {https://ui.adsabs.harvard.edu/abs/2004cxo..prop.1673W}
}

@ARTICLE{2018MNRAS.479.3634M,
       author = {{Marino}, Alessio and {Degenaar}, N. and {Di Salvo}, T. and {Wijnands}, R. and {Burderi}, L. and {Iaria}, R.},
        title = "{On obtaining neutron star mass and radius constraints from quiescent low-mass X-ray binaries in the Galactic plane}",
      journal = {\mnras},
         year = 2018,
        month = sep,
       volume = {479},
       number = {3},
        pages = {3634-3650},
          doi = {10.1093/mnras/sty1585},
archivePrefix = {arXiv},
       eprint = {1806.04557},
 primaryClass = {astro-ph.HE},
       adsurl = {https://ui.adsabs.harvard.edu/abs/2018MNRAS.479.3634M}
}

@article{wilms2000,
  author = {Wilms, J. and Allen, A. and McCray, R.},
  title = {On the absorption of {X}-rays in the interstellar medium},
  journal = {The Astrophysical Journal},
  volume = {542},
  number = {2},
  pages = {914--924},
  year = {2000},
  doi = {10.1086/317016}
}

@ARTICLE{2024ApJ...977L..17B,
       author = {{Bhattacharya}, Sayantan and {Bhattacharyya}, Sudip and {Shaw}, Gargi},
        title = "{XMM-Newton High-resolution Spectroscopy of EXO 0748─676 after Its Reemergence from a Long Quiescence}",
      journal = {\apjl},
         year = 2024,
        month = dec,
       volume = {977},
       number = {1},
          eid = {L17},
        pages = {L17},
          doi = {10.3847/2041-8213/ad9337},
archivePrefix = {arXiv},
       eprint = {2408.02715},
 primaryClass = {astro-ph.HE},
       adsurl = {https://ui.adsabs.harvard.edu/abs/2024ApJ...977L..17B}
}

@ARTICLE{1986ApJ...308..213G,
       author = {{Gottwald}, M. and {Haberl}, F. and {Parmar}, A.~N. and {White}, N.~E.},
        title = "{The Bursting Behavior of the Transient X-Ray Burst Source EXO 0748-676: A Dependence between the X-Ray Burst Properties and the Strength of the Persistent Emission}",
      journal = {\apj},
         year = 1986,
        month = sep,
       volume = {308},
        pages = {213},
          doi = {10.1086/164491},
       adsurl = {https://ui.adsabs.harvard.edu/abs/1986ApJ...308..213G}
}

@INPROCEEDINGS{2006SPIE.6270E..1VF,
       author = {{Fruscione}, Antonella and {McDowell}, Jonathan C. and {Allen}, Glenn E. and {Brickhouse}, Nancy S. and {Burke}, Douglas J. and {Davis}, John E. and {Durham}, Nick and {Elvis}, Martin and {Galle}, Elizabeth C. and {Harris}, Daniel E. and {Huenemoerder}, David P. and {Houck}, John C. and {Ishibashi}, Bish and {Karovska}, Margarita and {Nicastro}, Fabrizio and {Noble}, Michael S. and {Nowak}, Michael A. and {Primini}, Frank A. and {Siemiginowska}, Aneta and {Smith}, Randall K. and {Wise}, Michael},
        title = "{CIAO: Chandra's data analysis system}",
    booktitle = {Observatory Operations: Strategies, Processes, and Systems},
         year = 2006,
       editor = {{Silva}, David R. and {Doxsey}, Rodger E.},
       series = {Society of Photo-Optical Instrumentation Engineers (SPIE) Conference Series},
       volume = {6270},
        month = jun,
          eid = {62701V},
        pages = {62701V},
          doi = {10.1117/12.671760},
       adsurl = {https://ui.adsabs.harvard.edu/abs/2006SPIE.6270E..1VF}
}

@INPROCEEDINGS{1993AAS...183.5515H,
       author = {{Hertz}, P. and {Wood}, K.~S. and {Cominsky}, L.~R.},
        title = "{The Orbital Period of EXO0748-676: Secular Change or Stochastic Jitter?}",
    booktitle = {American Astronomical Society Meeting Abstracts},
         year = 1993,
       series = {American Astronomical Society Meeting Abstracts},
       volume = {183},
        month = dec,
          eid = {55.15},
        pages = {55.15},
       adsurl = {https://ui.adsabs.harvard.edu/abs/1993AAS...183.5515H}
}

@ARTICLE{2006ApJ...644.1090H,
       author = {{Heinke}, Craig O. and {Rybicki}, George B. and {Narayan}, Ramesh and {Grindlay}, Jonathan E.},
        title = "{A Hydrogen Atmosphere Spectral Model Applied to the Neutron Star X7 in the Globular Cluster 47 Tucanae}",
      journal = {\apj},
         year = 2006,
        month = jun,
       volume = {644},
       number = {2},
        pages = {1090-1103},
          doi = {10.1086/503701},
archivePrefix = {arXiv},
       eprint = {astro-ph/0506563},
 primaryClass = {astro-ph},
       adsurl = {https://ui.adsabs.harvard.edu/abs/2006ApJ...644.1090H}
}

@ARTICLE{2008MNRAS.387..268G,
       author = {{Galloway}, D.~K. and {{\"O}zel}, F. and {Psaltis}, D.},
        title = "{Biases for neutron star mass, radius and distance measurements from Eddington-limited X-ray bursts}",
      journal = {\mnras},
         year = 2008,
        month = jun,
       volume = {387},
       number = {1},
        pages = {268-272},
          doi = {10.1111/j.1365-2966.2008.13219.x},
archivePrefix = {arXiv},
       eprint = {0712.0412},
 primaryClass = {astro-ph},
       adsurl = {https://ui.adsabs.harvard.edu/abs/2008MNRAS.387..268G}
}

@ARTICLE{2002Natur.420...51C,
       author = {{Cottam}, J. and {Paerels}, F. and {Mendez}, M.},
        title = "{Gravitationally redshifted absorption lines in the X-ray burst spectra of a neutron star}",
      journal = {\nat},
         year = 2002,
        month = nov,
       volume = {420},
       number = {6911},
        pages = {51-54},
          doi = {10.1038/nature01159},
archivePrefix = {arXiv},
       eprint = {astro-ph/0211126},
 primaryClass = {astro-ph},
       adsurl = {https://ui.adsabs.harvard.edu/abs/2002Natur.420...51C}
}

@ARTICLE{2006Natur.441.1115O,
       author = {{{\"O}zel}, F.},
        title = "{Soft equations of state for neutron-star matter ruled out by EXO 0748 - 676}",
      journal = {\nat},
         year = 2006,
        month = jun,
       volume = {441},
       number = {7097},
        pages = {1115-1117},
          doi = {10.1038/nature04858},
archivePrefix = {arXiv},
       eprint = {astro-ph/0605106},
 primaryClass = {astro-ph},
       adsurl = {https://ui.adsabs.harvard.edu/abs/2006Natur.441.1115O}
}

@ARTICLE{2011MNRAS.414.1077Z,
       author = {{Zhang}, Guobao and {M{\'e}ndez}, Mariano and {Jonker}, Peter and {Hiemstra}, Beike},
        title = "{The distance and internal composition of the neutron star in EXO 0748-676 with XMM-Newton}",
      journal = {\mnras},
         year = 2011,
        month = jun,
       volume = {414},
       number = {2},
        pages = {1077-1081},
          doi = {10.1111/j.1365-2966.2011.18443.x},
archivePrefix = {arXiv},
       eprint = {1007.0647},
 primaryClass = {astro-ph.HE},
       adsurl = {https://ui.adsabs.harvard.edu/abs/2011MNRAS.414.1077Z}
}

@ARTICLE{2009MNRAS.394L.136M,
       author = {{Mu{\~n}oz-Darias}, T. and {Casares}, J. and {O'Brien}, K. and {Steeghs}, D. and {Mart{\'\i}nez-Pais}, I.~G. and {Cornelisse}, R. and {Charles}, P.~A.},
        title = "{Dynamical constraints on the neutron star mass in EXO 0748-676}",
      journal = {\mnras},
         year = 2009,
        month = mar,
       volume = {394},
       number = {1},
        pages = {L136-L140},
          doi = {10.1111/j.1745-3933.2009.00630.x},
archivePrefix = {arXiv},
       eprint = {0901.4119},
 primaryClass = {astro-ph.HE},
       adsurl = {https://ui.adsabs.harvard.edu/abs/2009MNRAS.394L.136M}
}

@ARTICLE{2011AA...528A.150D,
       author = {{D{\'\i}az Trigo}, M. and {Boirin}, L. and {Costantini}, E. and {M{\'e}ndez}, M. and {Parmar}, A.},
        title = "{XMM-Newton observations of the low-mass X-ray binary EXO 0748-676 in quiescence}",
      journal = {\aap},
         year = 2011,
        month = apr,
       volume = {528},
          eid = {A150},
        pages = {A150},
          doi = {10.1051/0004-6361/201016200},
archivePrefix = {arXiv},
       eprint = {1102.2640},
 primaryClass = {astro-ph.HE},
       adsurl = {https://ui.adsabs.harvard.edu/abs/2011A&A...528A.150D}
}

@ARTICLE{2012MNRAS.420...75R,
       author = {{Ratti}, E.~M. and {Steeghs}, D.~T.~H. and {Jonker}, P.~G. and {Torres}, M.~A.~P. and {Bassa}, C.~G. and {Verbunt}, F.},
        title = "{Optical spectroscopy of the quiescent counterpart to EXO 0748-676: a black widow scenario?}",
      journal = {\mnras},
         year = 2012,
        month = feb,
       volume = {420},
       number = {1},
        pages = {75-83},
          doi = {10.1111/j.1365-2966.2011.19999.x},
archivePrefix = {arXiv},
       eprint = {1110.3963},
 primaryClass = {astro-ph.HE},
       adsurl = {https://ui.adsabs.harvard.edu/abs/2012MNRAS.420...75R}
}

@ARTICLE{2017MNRAS.471.2605C,
       author = {{Cheng}, Zheng and {M{\'e}ndez}, Mariano and {D{\'\i}az-Trigo}, Mar{\'\i}a and {Costantini}, Elisa},
        title = "{The cooling, mass and radius of the neutron star in EXO 0748-676 in quiescence with XMM-Newton}",
      journal = {\mnras},
         year = 2017,
        month = nov,
       volume = {471},
       number = {3},
        pages = {2605-2615},
          doi = {10.1093/mnras/stx1452},
archivePrefix = {arXiv},
       eprint = {1706.04784},
 primaryClass = {astro-ph.HE},
       adsurl = {https://ui.adsabs.harvard.edu/abs/2017MNRAS.471.2605C}
}

@ARTICLE{2014ApJ...791...47D,
       author = {{Degenaar}, N. and {Medin}, Z. and {Cumming}, A. and {Wijnands}, R. and {Wolff}, M.~T. and {Cackett}, E.~M. and {Miller}, J.~M. and {Jonker}, P.~G. and {Homan}, J. and {Brown}, E.~F.},
        title = "{Probing the Crust of the Neutron Star in EXO 0748-676}",
      journal = {\apj},
         year = 2014,
        month = aug,
       volume = {791},
       number = {1},
          eid = {47},
        pages = {47},
          doi = {10.1088/0004-637X/791/1/47},
archivePrefix = {arXiv},
       eprint = {1403.2385},
 primaryClass = {astro-ph.HE},
       adsurl = {https://ui.adsabs.harvard.edu/abs/2014ApJ...791...47D}
}

@ARTICLE{2022MNRAS.510.4736K,
       author = {{Knight}, Amy H. and {Ingram}, Adam and {Middleton}, Matthew and {Drake}, Jeremy},
        title = "{Eclipse mapping of EXO 0748-676: evidence for a massive neutron star}",
      journal = {\mnras},
         year = 2022,
        month = mar,
       volume = {510},
       number = {4},
        pages = {4736-4756},
          doi = {10.1093/mnras/stab3722},
archivePrefix = {arXiv},
       eprint = {2201.02188},
 primaryClass = {astro-ph.HE},
       adsurl = {https://ui.adsabs.harvard.edu/abs/2022MNRAS.510.4736K}
}

@ARTICLE{2024PhRvD.109l3005M,
       author = {{Miao}, Zhiqiang and {Qi}, Liqiang and {Zhang}, Juan and {Li}, Ang and {Ge}, Mingyu},
        title = "{Thermal x-ray studies of neutron stars and the equation of state}",
      journal = {\prd},
         year = 2024,
        month = jun,
       volume = {109},
       number = {12},
          eid = {123005},
        pages = {123005},
          doi = {10.1103/PhysRevD.109.123005},
archivePrefix = {arXiv},
       eprint = {2402.02799},
 primaryClass = {astro-ph.HE},
       adsurl = {https://ui.adsabs.harvard.edu/abs/2024PhRvD.109l3005M}
}

@ARTICLE{2019ApJ...887L..21R,
       author = {{Riley}, T.~E. and {Watts}, A.~L. and {Bogdanov}, S. and others},
        title = "{A NICER View of PSR J0030+0451: Millisecond Pulsar Parameter Estimation}",
      journal = {\apjl},
         year = 2019,
        month = dec,
       volume = {887},
       number = {1},
          eid = {L21},
        pages = {L21},
          doi = {10.3847/2041-8213/ab481c},
archivePrefix = {arXiv},
       eprint = {1912.05702},
 primaryClass = {astro-ph.HE},
       adsurl = {https://ui.adsabs.harvard.edu/abs/2019ApJ...887L..21R}
}

@ARTICLE{2019ApJ...887L..24M,
       author = {{Miller}, M.~C. and {Lamb}, F.~K. and {Dittmann}, A.~J. and others},
        title = "{PSR J0030+0451 Mass and Radius from NICER Data and Implications for the Properties of Neutron Star Matter}",
      journal = {\apjl},
         year = 2019,
        month = dec,
       volume = {887},
       number = {1},
          eid = {L24},
        pages = {L24},
          doi = {10.3847/2041-8213/ab50c5},
archivePrefix = {arXiv},
       eprint = {1912.05705},
 primaryClass = {astro-ph.HE},
       adsurl = {https://ui.adsabs.harvard.edu/abs/2019ApJ...887L..24M}
}

@ARTICLE{2021ApJ...918L..28M,
       author = {{Miller}, M.~C. and {Lamb}, F.~K. and {Dittmann}, A.~J. and {Bogdanov}, S. and others},
        title = "{The Radius of PSR J0740+6620 from NICER and XMM-Newton Data}",
      journal = {\apjl},
         year = 2021,
        month = sep,
       volume = {918},
       number = {2},
          eid = {L28},
        pages = {L28},
          doi = {10.3847/2041-8213/ac089b},
archivePrefix = {arXiv},
       eprint = {2105.06979},
 primaryClass = {astro-ph.HE},
       adsurl = {https://ui.adsabs.harvard.edu/abs/2021ApJ...918L..28M}
}

@ARTICLE{2021ApJ...918L..27R,
       author = {{Riley}, Thomas E. and {Watts}, Anna L. and {Ray}, Paul S. and others},
        title = "{A NICER View of the Massive Pulsar PSR J0740+6620 Informed by Radio Timing and XMM-Newton Spectroscopy}",
      journal = {\apjl},
         year = 2021,
        month = sep,
       volume = {918},
       number = {2},
          eid = {L27},
        pages = {L27},
          doi = {10.3847/2041-8213/ac0a81},
archivePrefix = {arXiv},
       eprint = {2105.06980},
 primaryClass = {astro-ph.HE},
       adsurl = {https://ui.adsabs.harvard.edu/abs/2021ApJ...918L..27R}
}

@ARTICLE{2009MNRAS.396L..26D,
       author = {{Degenaar}, N. and {Wijnands}, R. and {Wolff}, M.~T. and {Ray}, P.~S. and {Wood}, K.~S. and {Homan}, J. and {Lewin}, W.~H.~G. and {Jonker}, P.~G. and {Cackett}, E.~M. and {Miller}, J.~M. and {Brown}, E.~F.},
        title = "{Chandra and Swift observations of the quasi-persistent neutron star transient EXO 0748-676 back to quiescence}",
      journal = {\mnras},
         year = 2009,
        month = jun,
       volume = {396},
       number = {1},
        pages = {L26-L30},
          doi = {10.1111/j.1745-3933.2009.00655.x},
archivePrefix = {arXiv},
       eprint = {0811.4582},
 primaryClass = {astro-ph},
       adsurl = {https://ui.adsabs.harvard.edu/abs/2009MNRAS.396L..26D}
}

@ARTICLE{2025ATel17191....1D,
       author = {{Degenaar}, N. and {Homan}, J. and {Cackett}, E. and {Wijnands}, R. and {Wolff}, M.~T. and {Buisson}, D.~J.~K.},
        title = "{The neutron star LMXB EXO0748-676 has returned to quiescence}",
      journal = {The Astronomer's Telegram},
         year = 2025,
        month = may,
       volume = {17191},
        pages = {1},
       adsurl = {https://ui.adsabs.harvard.edu/abs/2025ATel17191....1D}
}

@ARTICLE{2026JHEAp..5100535A,
       author = {{Aromal}, P. and {Kashyap}, Unnati and {Chakraborty}, Manoneeta and {Bhattacharyya}, Sudip and {Maccarone}, Thomas J. and {Choudhary}, Vijay},
        title = "{The 2024 outburst of the neutron star LMXB EXO 0748─676: An investigation of bursts and eclipses with astrosat}",
      journal = {Journal of High Energy Astrophysics},
         year = 2026,
        month = mar,
       volume = {51},
          eid = {100535},
        pages = {100535},
          doi = {10.1016/j.jheap.2025.100535},
       adsurl = {https://ui.adsabs.harvard.edu/abs/2026JHEAp..5100535A}
}

@ARTICLE{2020A&A...638L...2P,
       author = {{Parikh}, A.~S. and {Wijnands}, R. and {Homan}, J. and {Degenaar}, N. and {Wolvers}, B. and {Ootes}, L.~S. and {Page}, D.},
        title = "{Unexpected late-time temperature increase observed in the two neutron star crust-cooling sources XTE J1701-462 and EXO 0748-676}",
      journal = {\aap},
         year = 2020,
        month = jun,
       volume = {638},
          eid = {L2},
        pages = {L2},
          doi = {10.1051/0004-6361/202038198},
archivePrefix = {arXiv},
       eprint = {2005.12014},
 primaryClass = {astro-ph.HE},
       adsurl = {https://ui.adsabs.harvard.edu/abs/2020A&A...638L...2P}
}

@ARTICLE{1996ApJ...465..487V,
       author = {{Verner}, D.~A. and {Ferland}, G.~J. and {Korista}, K.~T. and {Yakovlev}, D.~G.},
        title = "{Atomic Data for Astrophysics. II. New Analytic Fits for Photoionization Cross Sections of Atoms and Ions}",
      journal = {\apj},
         year = 1996,
        month = jul,
       volume = {465},
        pages = {487},
          doi = {10.1086/177435},
archivePrefix = {arXiv},
       eprint = {astro-ph/9601009},
 primaryClass = {astro-ph},
       adsurl = {https://ui.adsabs.harvard.edu/abs/1996ApJ...465..487V}
}

@ARTICLE{2010ApJ...711L.148G,
       author = {{Galloway}, Duncan K. and {Lin}, Jinrong and {Chakrabarty}, Deepto and {Hartman}, Jacob M.},
        title = "{Discovery of a 552 Hz Burst Oscillation in the Low-Mass X-Ray Binary EXO 0748-676}",
      journal = {\apjl},
         year = 2010,
        month = mar,
       volume = {711},
       number = {2},
        pages = {L148-L151},
          doi = {10.1088/2041-8205/711/2/L148},
archivePrefix = {arXiv},
       eprint = {0910.5546},
 primaryClass = {astro-ph.HE},
       adsurl = {https://ui.adsabs.harvard.edu/abs/2010ApJ...711L.148G}
}

@ARTICLE{1985IAUC.4039....1P,
       author = {{Parmar}, A.~N. and {White}, N.~E. and {Giommi}, P. and {Haberl}, F. and {Pedersen}, H. and {Mayor}, M.},
        title = "{EXO 0748-676}",
      journal = {\iaucirc},
         year = 1985,
        month = feb,
       volume = {4039},
        pages = {1},
       adsurl = {https://ui.adsabs.harvard.edu/abs/1985IAUC.4039....1P}
}

@ARTICLE{2005ApJ...632.1099W,
       author = {{Wolff}, Michael T. and {Becker}, Peter A. and {Ray}, Paul S. and {Wood}, Kent S.},
        title = "{A Strong X-Ray Burst from the Low-Mass X-Ray Binary EXO 0748-676}",
      journal = {\apj},
         year = 2005,
        month = oct,
       volume = {632},
       number = {2},
        pages = {1099-1103},
          doi = {10.1086/444348},
archivePrefix = {arXiv},
       eprint = {astro-ph/0506515},
 primaryClass = {astro-ph},
       adsurl = {https://ui.adsabs.harvard.edu/abs/2005ApJ...632.1099W}
}

@ARTICLE{2011MNRAS.412.1409D,
       author = {{Degenaar}, N. and {Wolff}, M.~T. and {Ray}, P.~S. and {Wood}, K.~S. and {Homan}, J. and {Lewin}, W.~H.~G. and {Jonker}, P.~G. and {Cackett}, E.~M. and {Miller}, J.~M. and {Brown}, E.~F. and {Wijnands}, R.},
        title = "{Further X-ray observations of EXO 0748-676 in quiescence: evidence for a cooling neutron star crust}",
      journal = {\mnras},
         year = 2011,
        month = apr,
       volume = {412},
       number = {3},
        pages = {1409-1418},
          doi = {10.1111/j.1365-2966.2010.17562.x},
archivePrefix = {arXiv},
       eprint = {1007.0247},
 primaryClass = {astro-ph.HE},
       adsurl = {https://ui.adsabs.harvard.edu/abs/2011MNRAS.412.1409D}
}

@ARTICLE{2026JHEAp..5300595S,
       author = {{Subba}, Nirpat},
        title = "{High-inclination accretion in EXO 0748─676 during reactivation: Eclipse timing, burst behaviour, and a comptonized continuum}",
      journal = {Journal of High Energy Astrophysics},
         year = 2026,
        month = jul,
       volume = {53},
          eid = {100595},
        pages = {100595},
          doi = {10.1016/j.jheap.2026.100595},
       adsurl = {https://ui.adsabs.harvard.edu/abs/2026JHEAp..5300595S}
}

@ARTICLE{2012arXiv1201.5602P,
       author = {{Page}, Dany and {Reddy}, Sanjay},
        title = "{Thermal and transport properties of the neutron star inner crust}",
      journal = {arXiv e-prints},
         year = 2012,
        month = jan,
          eid = {arXiv:1201.5602},
        pages = {arXiv:1201.5602},
          doi = {10.48550/arXiv.1201.5602},
archivePrefix = {arXiv},
       eprint = {1201.5602},
 primaryClass = {nucl-th},
       adsurl = {https://ui.adsabs.harvard.edu/abs/2012arXiv1201.5602P}
}

@ARTICLE{2004ApJ...614L.121V,
       author = {{Villarreal}, Adam R. and {Strohmayer}, Tod E.},
        title = "{Discovery of the Neutron Star Spin Frequency in EXO 0748-676}",
      journal = {\apjl},
         year = 2004,
        month = oct,
       volume = {614},
       number = {2},
        pages = {L121-L124},
          doi = {10.1086/425737},
archivePrefix = {arXiv},
       eprint = {astro-ph/0409384},
 primaryClass = {astro-ph},
       adsurl = {https://ui.adsabs.harvard.edu/abs/2004ApJ...614L.121V}
}

@ARTICLE{2024GCN.36653....1D,
       author = {{D'Elia}, V. and {Kennea}, J.~A. and {Page}, K.~L. and {Parsotan}, T.~M. and {Neil Gehrels Swift Observatory Team}},
        title = "{Trigger 1236151: Swift detection of EXO 0748-676}",
      journal = {GRB Coordinates Network},
         year = 2024,
        month = jun,
       volume = {36653},
        pages = {1},
       adsurl = {https://ui.adsabs.harvard.edu/abs/2024GCN.36653....1D}
}

@INPROCEEDINGS{1996ASPC..101...17A,
       author = {{Arnaud}, K.~A.},
        title = "{XSPEC: The First Ten Years}",
    booktitle = {Astronomical Data Analysis Software and Systems V},
         year = 1996,
       editor = {{Jacoby}, George H. and {Barnes}, Jeannette},
       series = {Astronomical Society of the Pacific Conference Series},
       volume = {101},
        month = jan,
        pages = {17},
       adsurl = {https://ui.adsabs.harvard.edu/abs/1996ASPC..101...17A}
}

\end{document}